\documentclass[pdflatex,sn-basic]{sn-jnl}%

\usepackage{graphicx}%
\graphicspath{{figures/}}%
\usepackage{subcaption}%
\usepackage{array}%
\usepackage{multirow}%
\usepackage{amsmath,amssymb,amsfonts}%
\usepackage{amsthm}%
\usepackage{mathrsfs}%
\usepackage[title]{appendix}%
\usepackage{xcolor}%
\usepackage{textcomp}%
\usepackage{manyfoot}%
\usepackage{booktabs}%
\usepackage{tikz}%
\usetikzlibrary{arrows.meta,positioning}%
\definecolor{okBlue}{HTML}{0072B2}%
\definecolor{okOrange}{HTML}{E69F00}%
\definecolor{okGreen}{HTML}{009E73}%
\definecolor{okVermillion}{HTML}{D55E00}%
\definecolor{okInk}{HTML}{333333}%
\usepackage{siunitx}%
\DeclareSIUnit\week{weeks}%
\usepackage{url}%

\theoremstyle{thmstyleone}%

\theoremstyle{thmstyletwo}%

\theoremstyle{thmstylethree}%

\begin{document}

\title[Disruption Simulation on a Product-Annotated Supply Network]{Agent-Based Disruption Simulation on an Edge-Level Product-Annotated Supply Network of the Automotive Industry}

\author*[1]{\fnm{Benjamin} \sur{Rolf}}\email{benjamin.rolf@ovgu.de}

\author[2,3]{\fnm{Hiroyasu} \sur{Inoue}}\email{inoue@gsis.u-hyogo.ac.jp}

\author[1,4]{\fnm{Sebastian} \sur{Lang}}\email{sebastian.lang@ovgu.de}

\affil*[1]{\orgdiv{Institute for Engineering of Products and Systems}, \orgname{Otto-von-Guericke-University}, \orgaddress{\city{Magdeburg}, \country{Germany}}}

\affil[2]{\orgdiv{Graduate School of Information Science}, \orgname{University of Hyogo}, \orgaddress{\city{Kobe}, \country{Japan}}}

\affil[3]{\orgdiv{Center for Computational Science}, \orgname{RIKEN}, \orgaddress{\city{Kobe}, \country{Japan}}}

\affil[4]{\orgname{Fraunhofer Institute for Factory Operation and Automation IFF}, \orgaddress{\city{Magdeburg}, \country{Germany}}}

\abstract{Disruptions spread through supply networks when customers cannot replace the products they lose. We ask how portfolio-overlap and product-specific rules for selecting candidate suppliers change cascades when customers search for new suppliers and form new supply edges. From the MarkLines database, we construct a product-annotated network of about \num{77000} firms and \num{197000} supplier--customer edges in the global automotive industry. We extend an agent-based disruption model with edge-level products, a product-specific substitution rule based on the semantic similarity of product names, and calibrated capacities and inventories. Only \qty{6}{\percent} of the edges carry observed products. We impute the others and test the imputation against the annotated edges. We run \num{119000} simulations. With all other mechanics fixed, the product-specific rule makes the cascades after a hub failure four to five times larger than the portfolio-overlap rule of the original model. Under alternative imputations and similarity thresholds, the factor ranges from 1.2 to 18. Within the simulations, substitutability also shapes which suppliers are critical. Network topology largely determines which firms a shock reaches, and substitutability determines how many of them fail. Of the 50 most damaging suppliers under random disruptions, topology alone identifies 20 and an index that adds product substitutability identifies 32. The index carries over without refitting to a regional scenario modelled on the 2011 Tohoku Earthquake. Supply network risk assessment should account for product-level substitutability and therefore needs product data on supply relationships.}


\maketitle

\section{Introduction}\label{sec:intro}

Supply chain disruptions have become a persistent challenge for industrial production. The 2011 Tohoku Earthquake disrupted supply chains in Japan and abroad \citep{boehm2019}. It reduced Japan's gross domestic product (GDP) growth in the following year by around half a percentage point \citep{carvalho2021}. Its indirect value-added losses are estimated to have been more than one hundred times larger than the direct ones \citep{inoue2019}. The COVID-19 pandemic disrupted supply chains worldwide. Supply chain resilience has moved from an operational to a strategic concern \citep{sheffi2001, christopher2004, ponomarov2009}. Disruptions seldom remain local. A single failure can propagate through the supply network and produce what \citet{ivanov2014} call the ripple effect. Quantifying this effect requires an accurate representation of the network and a credible model of how firms react when their suppliers fail.

A large body of work analyses supply networks as complex systems and simulates the propagation of disruptions with agent-based models. Many of these models reduce substitutability to topological proximity or to firm attributes such as size, tier, or location. Recent work brings products into propagation models. \citet{inoue2025} show on Japanese production networks that defining inputs by product instead of industry roughly doubles the simulated share of inactive firms. \citet{amico2024} model how distributors reroute substitute products through existing supply links. In both models, a customer can only fall back on suppliers it is already linked to. Agent-based models with supplier rewiring let customers search for new suppliers instead. The most prominent of them, the model of \citet{zhao2019}, attaches products to firms. It treats any overlap between the portfolios of two firms as evidence that one can replace the other. On a small network this approximation is reasonable. On an industry-spanning network such as the global automotive supply chain, it offers the median lost input about \num{31000} candidate substitutes. It thereby understates how often substitution fails. A stricter rule needs to know which products a firm can supply and which product flows along each supply edge. We therefore ask how portfolio-overlap and lost-product-specific candidate-selection rules change cascades under shared supplier-search and rewiring conditions.

We make four contributions. First, we construct a product-annotated supply network of the global automotive industry from the MarkLines database. It has roughly \num{77000} firms and \num{197000} directed supplier--customer edges and maps products to firms and, where available, to individual edges. Second, we extend the agent-based model of \citet{zhao2019} with edge-level product allocation, a stricter product-specific substitution rule, an empirically calibrated capacity model, parallel supplier requests, and multi-firm disruption seeds. Third, we quantify how portfolio-overlap and lost-product-specific candidate selection affect cascades under shared model settings. We complement this comparison with a hold-out test of the product imputation and with sensitivity analyses of the imputation and the similarity threshold. Fourth, we decompose the criticality of a supplier into its exposure, its downstream reach, and the substitutability of its products. Only substitutability requires product data. Within the simulations, adding it to a topological index of downstream reach improves the identification of the most damaging suppliers. This holds under random shocks and, without refitting, under a regional shock.

Section~\ref{sec:related} reviews related work on supply network science, firm-level disruption simulation, and resilience. Section~\ref{sec:data} describes the MarkLines data and the topology of the resulting network, and Section~\ref{sec:methods} presents the simulation model and the experimental design. Section~\ref{sec:results} reports three experimental campaigns. A baseline comparison isolates the substitution rule, a Monte Carlo analysis varies the shock magnitude, and a regional scenario follows the geography of the 2011 Tohoku Earthquake. Section~\ref{sec:discussion} discusses the results together with the strengths and limitations of the approach, and Section~\ref{sec:conclusions} concludes.

\section{Related work}\label{sec:related}

Three streams of literature are relevant to this paper. The first treats supply chains as complex networks and reconstructs them. The second covers agent-based simulation of cascading disruptions. The third is the conceptual literature on resilience and the ripple effect.

\subsection{Supply network science and network reconstruction}\label{sec:rw-ns}

Supply networks have long been studied as complex adaptive systems. \citet{borgatti2009} brought the vocabulary of social network analysis to supply chain management. \citet{bellamy2013} organised the field around network structure, network dynamics, and network strategy, and \citet{perera2017} review the topological evidence gathered since. Empirical work has moved from small, fully mapped networks to large, partially observed ones. \citet{choi2002} mapped by hand the supply networks behind the centre-console assembly of three vehicle lines. Manual mapping cannot reach this level of detail beyond individual assemblies. The Airbus aerospace network of \citet{brintrup2017} is an early example of large-scale secondary data. Its communities are connected by hub firms that make the system cohesive and, at the same time, vulnerable to targeted disruption. The MarkLines network shows a similar structure.

With more than 300 million firms connected by an estimated 13 billion supply links worldwide \citep{pichler2023}, the mapping of firm-level supply networks has become a research field of its own. It draws on administrative records such as the Hungarian value-added tax (VAT) network behind the systemic risk analysis of \citet{diem2022}. Other studies reconstruct networks from partial information with statistical and machine learning methods \citep{squartini2018, mungo2023}. Others extract supply links from news and other public texts \citep{wichmann2020}. Recent work uses large language models (LLMs) for this task \citep{almahri2026}. Commercial databases cover more firms than most public sources, but they are curated and incomplete. \citet{bacilieri2026} show that routinely reported topological properties are biased in datasets with missing data. MarkLines is such a dataset.

\subsection{Firm-level disruption simulation}\label{sec:rw-sim}

If the shocks hitting the $n$ sectors of an economy were independent, the volatility of aggregate output would decline in proportion to $1/\sqrt{n}$. \citet{acemoglu2012} showed that this diversification argument fails once sectors are connected through input--output linkages and a few of them supply far more of the economy than the rest. From network physics, \citet{albert2000} add that networks with heavy-tailed degree distributions tolerate random failures but collapse when their hubs are removed.

Agent-based models describe how firms react to a shock: they search for alternative suppliers and fail if they find none. \citet{nair2011} simulated inventory dynamics on \num{20} scale-free and random topologies. Their firms trade a single generic good and do not switch suppliers. Orders to a failed firm wait until it recovers. \citet{namdar2024} showed that modular structure limits the spread of cascades across module boundaries while it amplifies their severity within modules. \citet{zhao2019} introduced products, and their model is our baseline. On a \num{2971}-firm network built from Mergent Horizon data, a disrupted customer searches for substitutes among the competitors of its failed supplier. It weights each competitor by the overlap of its product portfolio with that of the failed supplier. MarkLines has no competitor lists. Our portfolio-overlap rule therefore treats every firm that shares a product with the failed supplier as a competitor and weights it in the same way.

Recent studies bring product granularity into propagation analyses. \citet{inoue2025} show on Japanese production networks that defining inputs by product instead of industry raises the simulated share of inactive firms from about \qty{11}{\percent} to \qty{23}{\percent}. Their customers can only fall back on existing suppliers of the same product. \citet{amico2024} show for the US opioid distribution system that rerouting substitute products through existing supply links reduces and delays shortages. \citet{berthou2024} combine product-level trade data with input--output tables to map product-level vulnerabilities of country-industries. Their partial-equilibrium model covers only the first tier of exposure. None of these studies lets a customer search for new suppliers and form new edges.

The strongest empirical evidence that substitution constraints drive propagation comes from the 2011 Tohoku Earthquake. \citet{boehm2019} estimate a near-zero elasticity of substitution between imported and domestic inputs at US affiliates of Japanese multinationals. \citet{inoue2019} simulate the shock across nearly one million Japanese firms and find indirect value-added losses more than one hundred times larger than the direct ones. \citet{carvalho2021} estimate that the earthquake lowered Japan's real GDP growth in the following year by 0.47 percentage points. Without input--output linkages between firms inside and outside the disaster area, the decline would have been 0.21 percentage points. For three decades of natural disasters in the United States, \citet{barrot2016} show that customer losses concentrate where intermediate inputs are specific. \citet{diem2022} find that the default of any one of 32 firms ($\qty{0.035}{\percent}$ of all firms) would reduce Hungarian production by about $\qty{23}{\percent}$. \citet{son2021} document how the supply networks of the affected Japanese firms changed after the earthquake.

\subsection{Resilience concepts and the ripple effect}\label{sec:rw-resilience}

Disruption studies borrow their vocabulary from the management literature. \citet{sheffi2001} called for preparing for disruptions through planned redundancies, \citet{christopher2004} and \citet{ponomarov2009} defined resilience, and \citet{wieland2021} reframed it as adaptation and transformation. Two constructs describe the threat side. The ripple effect of \citet{ivanov2014} denotes disruptions that propagate through the supply network and change its structure as supply relationships are lost and re-established. The criticality of the nexus supplier of \citet{yan2015} stems from its position in the surrounding network and not from its internal capabilities. Buying firms therefore often recognise such a supplier only once a disruption occurs. So far, the construct has been developed conceptually and tested in agent-based simulations on real-world supply networks of a few hundred firms each \citep{namdar2024}. Section~\ref{sec:res-exp-C} examines it within the simulation on an empirical, industrial-scale network.

The closest studies differ from ours in how a customer replaces a lost input. In \citet{inoue2025} and \citet{amico2024}, products determine which existing suppliers can step in. In \citet{zhao2019}, customers search for new suppliers, but firm portfolios determine the candidates. \citet{berthou2024} work with country-industry aggregates and do not model this search. We are not aware of a study that compares portfolio-overlap and lost-product-specific candidate selection under the same supplier search and rewiring. Such a comparison requires an industry-scale firm network with products on its edges. \citet{brintrup2017} name edge-level product information as missing from supply network data. We build such a network from MarkLines.

\section{Data}\label{sec:data}

\subsection{Data source}\label{sec:data-source}

Our work is based on MarkLines \citep{marklinescoltd2026}, a commercial database of production, sales, and supplier information for the global automotive industry. The database covers the supply chain from original equipment manufacturers (OEMs) through Tier-1 system suppliers to Tier-$n$ component manufacturers and material providers. Compared with the supply network datasets commonly used in academic work, it has two advantages. First, it includes privately held firms and therefore reaches beyond disclosure-driven datasets such as Mergent Horizon or FactSet. Second, it maps each firm to the products it can supply. For part of the supply relations, mostly those between Tier-1 suppliers and OEMs, MarkLines also names the delivered products. Our edge-level extension of the \citet{zhao2019} baseline builds on this information. MarkLines has been used before to map the Toyota supply network \citep{kito2014, arora2021} and to predict supply links \citep{zheng2025}. These studies attach products, if at all, to firms and not to individual supply edges.

As a curated industry database, MarkLines is neither an administrative record nor an exhaustive ground truth, and we see it as a complement to the mapping efforts surveyed by \citet{pichler2023}. Its supply information is biased towards Tier-1-to-OEM relations. This bias exaggerates the flatness of the tier hierarchy and the prevalence of Tier-1 hubs. Moreover, only a small fraction of its edges carry product annotations. We impute the missing products (see Section~\ref{sec:mc-imputation}) instead of dropping the unannotated edges. The first caveat is a coverage bias of the kind that \citet{bacilieri2026} document for commercial datasets. Such datasets miss many firms and relationships and over-represent large firms.

\subsection{Resulting network}\label{sec:network-stats}

MarkLines mixes plant-level entities (e.g.\ Toyota Motor Kyushu, Miyata Plant) with their parent companies and splits its information across unconnected tables. We therefore consolidate the records into firm groups and match them across tables (see Appendix~\ref{app:data} and Figure~\ref{fig:pipeline}). The resulting network contains \num{76986} firms and \num{197077} directed supplier--customer edges, and \num{1201} firms are flagged as OEMs. The largest weakly connected component (LCC) contains \num{32668} firms, and \num{44244} firms are isolated because MarkLines covers none of their supply relations. The product vocabulary $P$ comprises \num{41098} product names. We embed them jointly so that the cosine similarity of two embeddings can approximate the substitutability of the two products.

Product coverage is structurally asymmetric. At the node level, \num{70843} firms (\qty{92.0}{\percent}) carry at least one product attribute, the \num{1201} OEMs implicitly supply vehicles, and only \num{4942} firms (\qty{6.4}{\percent}) have neither. At the edge level, only \num{12411} edges (\qty{6.3}{\percent}) carry a product annotation. The annotation lists the delivered products together with the vehicle models they are supplied for. The number of product--model entries on an edge serves as a proxy for its supply volume. The remaining \num{184666} edges (\qty{93.7}{\percent}) only record that $u$ supplies $v$. As the product-specific rule requires a product label on every edge, we impute the missing labels.

\subsection{Network topology}\label{sec:topology}

\begin{figure}[tbp]
  \centering
  \includegraphics[width=\linewidth]{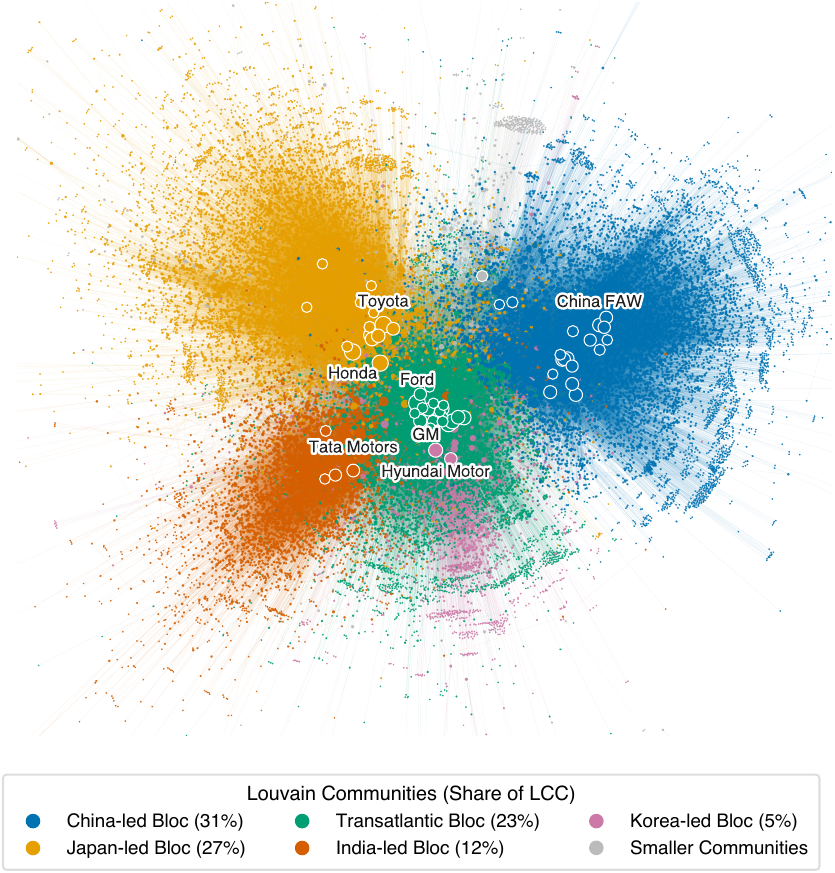}
  \caption{Overview of the MarkLines supply network, largest connected component (\num{32668} firms). Node size is proportional to degree. The highest-degree hubs are ringed in white. The highest-degree firm of each major community and the two remaining firms with the highest degree are labelled. Colours mark the five largest Louvain communities. The legend names them after their anchor economies. Smaller communities are shown in grey. Intra-community edges are tinted in the community colour. Inter-community edges are grey.}
  \label{fig:network-overview}
\end{figure}

The total, in-, and out-degree distributions are heavy-tailed. As in the FactSet data analysed by \citet{bacilieri2026}, the in-degree tail is heavier than the out-degree tail. Consistent with \citet{broido2019}, likelihood-ratio tests favour the power law only for total degree and the log-normal distribution for in- and out-degree. We therefore use ``heavy-tailed'' throughout and report the complementary cumulative distribution functions (CCDFs), tail exponents, and goodness-of-fit statistics in Appendix~\ref{app:topology}. Beyond the degree sequence, the network is a short-path, modular, near-acyclic system with a core--periphery layout. Clustering is weak (average clustering coefficient $C_\text{cl} = 0.028$). The small-world index is $\sigma = 1.70$. The tier hierarchy is shallow because automotive supply networks are flat and MarkLines covers little beyond Tier 1. Louvain partitioning \citep{blondel2008} yields 18 communities with modularity $Q = 0.5346$. The five largest cover approximately \qty{98}{\percent} of the LCC and correspond to the major regional blocs of the automotive industry (see Figure~\ref{fig:network-overview}). We refer to them as the China-led, Japan-led, transatlantic (Germany--US), India-led, and Korea-led blocs, after their anchor economies. Neither country nor tier alone reproduces this structure, and the transatlantic bloc is the largest external partner of each of the four other blocs. Degree assortativity is close to zero, and a bow-tie decomposition reveals a nearly directed acyclic graph (DAG). This structure allows a breadth-first search (BFS) to assign the tiers on which the trial budgets of the model depend.

Compared with Mergent Horizon, the $\num{2971}$-firm dataset on which \citet{zhao2019} developed and ran the baseline model, MarkLines is more than an order of magnitude larger and carries edge-level product attributes. On such a network, the portfolio-overlap rule of the baseline may overestimate substitutability and underpredict propagation.

\section{Methods}\label{sec:methods}

The simulation model extends the baseline of \citet{zhao2019} in four respects. It runs on the MarkLines network and replaces the portfolio-overlap rule with a product-specific substitution rule. It extends the search procedure and calibrates capacity and inventory parameters empirically. A Monte Carlo imputation supplies the missing edge products.

\subsection{Simulation model}\label{sec:model}

The simulation is an agent-based, discrete-time model of disruption propagation on a directed supply network. Each firm is an agent that observes its incoming supply edges, tries to replace a failed supplier with a substitute, and becomes disrupted itself if it does not succeed. Our baseline is the agent-based model of \citet{zhao2019}. One simulation step corresponds to one week. We assume that a business-to-business (B2B) request-for-quote (RFQ) cycle for industrial inputs takes about one week. A customer therefore makes one substitution attempt per step.

\begin{figure}[tbp]
  \centering
  \begin{subfigure}{\linewidth}
    \centering
    \includegraphics[width=\linewidth]{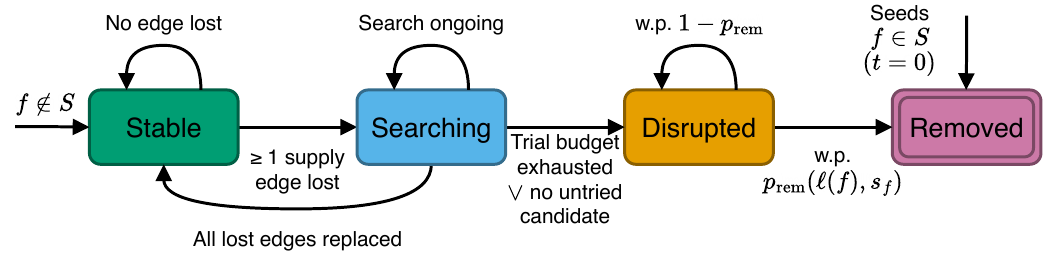}
    \subcaption{Reactive state machine of each firm agent. Non-seed firms start in Stable, while seed firms start in Removed. Losing a supply edge triggers the Searching state. A firm returns to Stable once all lost edges have been re-established. It becomes Disrupted when a lost edge has exhausted its trial budget or no untried candidate remains for a lost product. Disrupted firms are removed with a weekly probability $p_\text{rem}$ drawn around $\mu(f)$ (see Equation~\ref{eq:removal}). This probability increases with the lost share of inbound supply $\ell(f)$ and decreases with firm size $s_f$. Removed is absorbing.}\label{fig:model-states}
  \end{subfigure}\\[2ex]
  \begin{subfigure}{\linewidth}
    \centering
    \includegraphics[width=\linewidth]{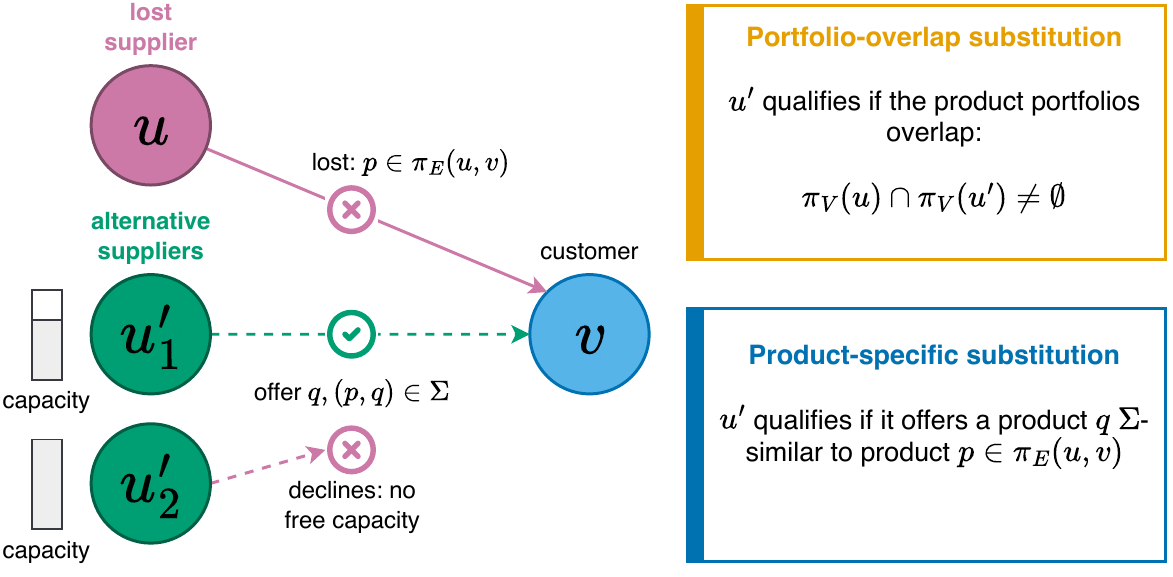}
    \subcaption{Substitution search after supplier $u$ fails on edge $(u,v)$. The portfolio-overlap rule allows any firm with a shared portfolio product as a substitute, whereas the product-specific rule requires the product to be similar to the lost products $\pi_E(u,v)$. Candidates accept requests only up to their free capacity and prioritise existing partners.}\label{fig:model-substitution}
  \end{subfigure}
  \caption{Model schematic.}
  \label{fig:model-schematic}
\end{figure}

The baseline of \citet{zhao2019} encodes the system through two networks. The supply network $G_s = (V, E_s)$ consists of the firm set $V$ and the supply edge set $E_s$. In the competition network $G_p$, the edge weights are the overlaps of the firms' product portfolios and rank the substitution candidates. Each firm is in one of four states: Stable, Searching, Disrupted, or Removed (see Figure~\ref{fig:model-states}). Each weekly cycle comprises two half-steps. In the first, searching firms send supply requests to potential suppliers. In the second, candidate suppliers accept or decline requests up to their free capacity. They serve existing partners first. They accept the remaining requests at random with a probability proportional to the size of the requesting firm. Each lost edge carries a trial budget, the number of substitution attempts the customer may make for that edge. A firm becomes Disrupted when a lost edge exhausts its budget or when no untried candidate remains for a lost input. At the end of each cycle, disrupted firms are removed stochastically. Following \citet{zhao2019}, the weekly removal probability of a disrupted firm $f$ is drawn around
\begin{equation}\label{eq:removal}
  \mu(f) = 1 - \frac{\ln s_f - \ln s_\text{min} + 1}{\ln s_\text{max} - \ln s_\text{min} + 1}\,\bigl(1 - \ell(f)\bigr),
\end{equation}
either from a normal distribution with standard deviation $0.1$ or uniformly within $\mu(f) \pm 0.1$ (see Table~\ref{tab:zhao-parameters}). Here $s_f$ is the size of $f$, and $s_\text{min}$ and $s_\text{max}$ are the smallest and largest firm sizes in the network. The loss ratio $\ell(f)$ is the share of $f$'s original inbound supply volume that is currently lost. Re-established supply lowers $\ell(f)$, and with the unit edge weights of the baseline, $\ell(f)$ is the share of lost suppliers. \citet{zhao2019} measure firm size as revenue. Because MarkLines does not report revenue for most privately held firms, we proxy size by the firm's degree in the supply network. A simulation run is initialised by marking a seed set $S$ of firms as Removed and runs for a fixed horizon (see Table~\ref{tab:zhao-parameters}). It does not stop when the last search has ended, because Disrupted firms can still be removed. Firms that are still Disrupted at the horizon are not counted as removed. On average, this concerns $0.5$ firms per run and never more than four. At the network level, the total supply capacity is $C(t) = \sum_{f \in V_t} c_\text{out}(f, t)$. Here $V_t$ is the set of firms still active at time $t$, and $c_\text{out}(f, t)$ is the sum of the supply volumes on the edges that leave $f$ at time $t$. The supply volume of an edge is the number of its product--model delivery entries. These entries are observed on annotated edges and imputed on the others. When a firm is removed, all of its supply edges vanish, so $C(t)$ drops by its outbound volume and by the volume its suppliers delivered to it. When a searching customer re-establishes a lost edge with a substitute supplier, the corresponding volume returns to $C(t)$. In this way, $C(t)$ measures the cumulative extent of a cascade in volume terms, and its weekly change $\Delta C(t) = C(t) - C(t-1)$ measures its pace. Both weight the loss of large firms more heavily than a plain count of removed firms.

Our model extends this baseline along four axes that together constitute the methodological contribution of this paper:
\begin{enumerate}
  \item \textbf{Dataset}: we replace the $\num{2971}$-firm Mergent Horizon network used by \citet{zhao2019} with the $\num{76986}$-firm MarkLines automotive supply network. This increases the number of firms and edges by more than an order of magnitude and adds product annotations to the edges.
  \item \textbf{Product-specific substitutability}: we replace the firm-level portfolio-overlap rule with the product-specific rule. This embedding-based criterion matches a lost product against semantically similar candidate products instead of the supplier's full portfolio. Edge-level input types are not new per se. For example, the adaptive regional input--output (ARIO) lineage of disaster models \citep{hallegatte2014} distinguishes inputs by the supplying sector. The firm-level simulation of \citet{inoue2019} allows replacement only among suppliers of the same sector. \citet{inoue2025} allow it among existing suppliers of the same product. In our model, the lost product determines which firms a customer can approach, including firms it is not yet linked to.
  \item \textbf{Extended simulation logic}: we extend the reactive search procedure along two dimensions. The original model restricts disruption initialisation to a single seed firm (e.g.\ a bankruptcy). The regional scenario of Experiment~3 requires multi-firm seed sets, and we allow seed sets of any size. The baseline also contacts only one substitution candidate per cycle. We replace this with $k$ parallel requests, as B2B buyers do in RFQ procurement: a searching customer contacts $k$ candidates per cycle and accepts the first to respond. The extended configuration uses $k = 3$ in all campaigns.
  \item \textbf{Empirical parameter calibration}: we replace two of the baseline's ad hoc parameter choices with industry-based distributions. In the baseline, a supplier accepts a limited number of requests. This limit is drawn from one of three normal distributions for the smallest, middle, and largest third of the suppliers by size and rounded to whole requests (see Table~\ref{tab:zhao-parameters}). We replace it by a spare capacity ratio $\theta$. Each supplier $f$ draws its own ratio $\theta_f$. Its free capacity starts at $c_\text{free}(f) = \theta_f\, c_\text{out}(f, 0)$ rounded to whole units. Current supply and free capacity together form the capacity of $f$. Its utilisation $1/(1 + \theta_f)$ has a mean of about $0.80$. This is close to the capacity utilisation of about \qty{78}{\percent} in European manufacturing in the Eurostat series \texttt{teibs070} \citep{eurostat2026}. Almost all suppliers with an outbound volume of one unit and about half of those with two units therefore start without free capacity. The free capacity decreases with every accepted request and increases by the volume of every customer the supplier loses. Firms without recorded customers have no observed outbound volume. A free capacity of zero would reflect the coverage of MarkLines rather than their production. They therefore receive a free capacity of one unit. The fixed trial budget for finding alternative suppliers is replaced by a days-of-supply representation of inventories, as in \citet{hallegatte2014}. The budget of an edge is the number of whole weeks in the sum of the customer's inbound and the supplier's outbound days of supply. Each firm draws its inbound and its outbound days of supply independently. Based on the illustrative automotive inventory ranges of \citet{alicke2020}, we use $\mathcal{N}(12, 5)$ for tiers 0--1 and $\mathcal{N}(35, 5)$ for deeper tiers. The budget caps the search time of an edge. It is not a guaranteed grace period. As in the baseline, a search also ends when no untried candidate remains. A customer can therefore become Disrupted before its inventory runs out.
\end{enumerate}
Table~\ref{tab:zhao-parameters} summarises the parameters of both configurations. All other baseline mechanics are adopted unchanged. Missing edge-product information is handled by Monte Carlo imputation.

\begin{table}[tbp]
  \centering
  \caption{Parameters of the baseline configuration of \citet{zhao2019} and of the extended configuration. The hybrid configuration of Experiment~1 uses the extended parameters with the portfolio-overlap rule. Here $\mathcal{N}(m, s)$ denotes a normal distribution with mean $m$ and standard deviation $s$, and $\mathcal{U}(a, b)$ a uniform distribution on $[a, b]$. Draws are truncated to $\kappa \geq 1$, $0 \leq \theta \leq 1$, and non-negative values for all other parameters. A removal probability above 1 means certain removal. Equation~\ref{eq:removal} gives the centre $\mu$ of the removal probability.}
  \label{tab:zhao-parameters}
  \small
  \begin{tabular}{@{}lll@{}}
    \toprule
    Parameter & Baseline & Extended \\
    \midrule
    Substitution rule & portfolio overlap & product-specific \\
    Similarity threshold $\tau$ & -- & $0.73$ \\
    Parallel requests per cycle $k$ & $1$ & $3$ \\
    Trial budget per lost edge & $10$ attempts & days of supply \\
    Acceptance limit per supplier & $\mathcal{N}(2, 0.5)$, $\mathcal{N}(4, 1)$, $\mathcal{N}(6, 2)$ & $\theta \sim \mathcal{N}(0.25, 0.1)$ \\
    Free capacity, no recorded customers & as above & $1$ unit \\
    Partner inertia $\kappa$ & $\mathcal{N}(1.5, \sigma_\kappa)$, $\sigma_\kappa \in \{0.1, 0.2\}$ & as baseline \\
    Weekly removal probability & $\mathcal{N}(\mu, 0.1)$ or $\mathcal{U}(\mu - 0.1, \mu + 0.1)$ & as baseline \\
    Simulation horizon & \qty{20}{\week} & \qty{52}{\week} \\
    \bottomrule
  \end{tabular}
\end{table}

\subsubsection{Edge-level product representation}\label{sec:edge-granularity}

The main methodological change in this paper is the representation of products. Most prior agent-based supply-network propagation models omit product information from the simulation altogether. \citet{zhao2019} associate products with firms via a node-level product map $\pi_V: V \to 2^P$, where $P$ is the global product vocabulary. They then use the pairwise product overlap between two firms to determine the edge weights in the competition network $G_p$. We retain $\pi_V$ but additionally introduce an edge-level product map $\pi_E: E_s \to 2^P$ that records, for each supply edge $(u, v)$, the specific product set delivered along that edge. On imputed edges, the delivered products are a subset of the supplier's portfolio by construction ($\pi_E(u, v) \subseteq \pi_V(u)$). Annotated edges may name products that the portfolio lists under a different name or not at all.

An example illustrates the difference. In MarkLines, the $|\pi_V(u)| = 101$ products in the portfolio of Robert Bosch Korea range from braking and engine management to infotainment and door modules. Its annotated supply edge to Beijing Hyundai carries $\pi_E(u, v) = \{\text{Engine ECU},\ \text{On-board diagnostics}\}$, two engine electronics items from this portfolio. The firm-level representation treats all 101 products as relevant for a failure on this edge. The edge-level representation ties the impact of a Bosch Korea failure on Beijing Hyundai to the substitutability of these two products.

The edge-product map underpins the substitution logic. When supplier $u$ fails to supply customer $v$, the customer searches for substitutes for the specific lost products $\pi_E(u, v)$ (see Figure~\ref{fig:model-substitution}). It does not search for firms that share some product with $u$. A hierarchical hash from products to candidate suppliers allows a constant-time lookup of the candidates per lost product. The change to the model definition is small, but its effect on the search is large. A customer no longer searches the firm-level overlap set but a per-product candidate set. This set has a median size of about 830 firms and ranges from no firm for niche products to about \num{21000} firms for widely produced parts (see Figure~\ref{fig:candidate-pools}).

\subsubsection{Substitution logic}\label{sec:substitution}

The substitutability criterion is a central design choice of any disruption model. It determines how easily customers can replace a failed supplier and thus whether local failures are absorbed or turn into cascades. The portfolio-overlap rule of the baseline accepts any supplier that shares at least one product with the failed supplier. In small networks such as the \num{2971}-firm Mergent Horizon dataset of \citet{zhao2019}, this test is selective enough to produce cascades. MarkLines is more than an order of magnitude larger and has broad product portfolios. There, the median product lost on a failed edge has about \num{31000} candidates under this rule (see Figure~\ref{fig:candidate-pools}). Experiment~1 tests how this permissiveness changes the cascades.

The embedding-based substitutability criterion that we adopt as our default operates at the level of products rather than firms:
\begin{enumerate}
    \item For each unique product $p$ in the product vocabulary $P$, we compute a dense \num{3072}-dimensional vector representation using the \texttt{text-embedding-3-large} model of OpenAI.
    \item Two products $p$ and $q$ count as mutually substitutable when their cosine similarity exceeds a threshold of $\tau = 0.73$. We chose this threshold by manually inspecting the false-positive and false-negative rates on a stratified inspection sample. Experiment~1 varies $\tau$ between $0.65$ and $0.85$. The resulting product--product substitutability map is a sparse, symmetric, and reflexive relation $\Sigma \subseteq P \times P$. A product's similarity to itself is $1$, so identical products always qualify as substitutes.
    \item If supplier $u$ fails on its supply edge $(u, v)$, customer $v$ searches substitutes for each lost product $p \in \pi_E(u, v)$. The candidates are the suppliers that can deliver a product $q$ similar to $p$, $\mathcal{C}(p) = \{u' \in V_t \setminus \{v\} : \exists\, q \in \pi_V(u') \cup \pi_D(u'),\ (p, q) \in \Sigma\}$. The candidates exclude the customer and all removed firms, including the failed supplier. Here $\pi_D(u')$ is the set of products that $u'$ is observed to deliver on annotated edges. A firm that demonstrably supplies a product can also supply it to other customers, even if its portfolio does not list it. Each candidate is scored by the highest similarity among its matching products. The customer approaches candidates at random with a probability proportional to this score. For existing partners, the partner-inertia factor $\kappa$ multiplies the score (see Table~\ref{tab:zhao-parameters}).
\end{enumerate}

The criterion is product-specific because the search starts from the lost products $\pi_E(u, v)$ and not from the supplier's full portfolio $\pi_V(u)$. It is semantic because $\Sigma$ derives from product-name embeddings instead of product overlap counts. The literature on input specificity motivates it empirically. \citet{barrot2016} show that customers of disrupted suppliers lose output and market value precisely when the intermediate inputs are specific. \citet{boehm2019} estimate from the 2011 Tohoku Earthquake that the elasticity of substitution between imported and domestic inputs is close to zero. This is consistent with a Leontief production function. Against this evidence, the portfolio-overlap rule is too permissive wherever inputs are specific. Experiment~1 tests the resulting hypothesis that it underestimates propagation on a large industry-level network.

Semantic similarity of product names can only approximate substitutability. Whether a firm can replace a lost input depends on specifications, qualification, and tooling that product names do not reveal. Two checks bound the errors of this proxy (see Appendix~\ref{app:validation}). Of the $\Sigma$ pairs of two different MarkLines standard products, \qty{16.5}{\percent} lie in different product families. This share is a lower bound for implausible matches. Where an OEM sources the same product for the same vehicle model and model year from two suppliers, each supplier is a revealed substitute for the other. The product-specific rule lists the second supplier as a candidate in \qty{86}{\percent} of these \num{3282} cases, and the portfolio-overlap rule in \qty{92}{\percent}. The product-specific rule achieves this with a median of 146 candidates per product instead of tens of thousands.

\subsubsection{Monte Carlo product imputation}\label{sec:mc-imputation}

The product-specific substitutability criterion requires a product label on every supply edge. Only $\qty{6.3}{\percent}$ of MarkLines edges carry an explicit product annotation $\pi_E(u, v)$. The remaining $\qty{93.7}{\percent}$ specify only that $u$ supplies $v$. We address this gap by a Monte Carlo imputation with three sampling steps per unannotated edge.
\begin{enumerate}
    \item The number of products carried by edge $(u,v)$ is drawn from a kernel density estimate (KDE) of the product count distribution observed on the annotated edges.
    \item The products supplied along edge $(u,v)$ are drawn uniformly from the supplier's product portfolio $\pi_V(u)$. In the sensitivity runs of Experiments~1 and~2, a product $p$ is instead drawn with probability proportional to $n_p^{\gamma}$. Here $n_p$ is the number of firms offering a product similar to $p$. A negative $\gamma$ favours products with few substitutes and $\gamma = 0$ is the uniform draw.
    \item The supply volume per product is drawn from a product-specific KDE of the volume distribution observed for that product on the annotated subset. Products that never appear on an annotated edge inherit the distribution of their most similar product.
\end{enumerate}

As the second sampling stage draws from the supplier's portfolio $\pi_V(u)$, the simulation excludes the \num{4261} firms without product information. The simulated network has \num{72725} firms and \num{197075} edges. OEMs and product-less pure customers are retained because product labels are required only on outgoing edges.

We tested the imputation on the \num{12411} annotated edges by hiding their products and re-imputing them from the supplier portfolios (see Appendix~\ref{app:validation}). Uniform draws understate how scarce the lost products are. The observed products have a median of 147 substitution candidates, and \qty{12.0}{\percent} of them have ten or fewer. The imputed products have a median of 623, and only \qty{5.4}{\percent} have ten or fewer. A tilt of $\gamma = -1$ reproduces the observed distribution best. We nevertheless keep the uniform draw as the main specification, because the annotated edges are not representative. Almost all of them (\qty{99}{\percent}) end at a firm without customers of its own, mostly an OEM, and only \qty{27}{\percent} of their product names occur in the portfolio vocabulary. Experiment~1 therefore also reports the substitution effect for $\gamma \in \{-2, -1, +1\}$, and Experiment~2 for $\gamma = -1$.

Experiment~2 uses 500 independent network configurations from this imputation procedure, and Experiment~3 uses 250. Every simulation result carries the uncertainty of the imputation.

\section{Results}\label{sec:results}

The model is implemented in \texttt{Python} 3.13.9. All campaigns ran on a SLURM-managed high-performance computing (HPC) cluster with fixed random seeds for the Monte Carlo imputation and the simulation. Given access to the MarkLines data, every reported result can therefore be reproduced with the published code and configuration \citep{Rolf2026.code}. At runtime, invariant assertions check the implementation, for example that no remaining capacity becomes negative and that outbound supply volume is conserved.

\subsection{Experiment~1: Baseline comparison}\label{sec:res-exp-A}

Experiment~1 isolates the effect of the substitution rule. The baseline and the extended configuration differ in four respects at once: the substitution rule, the capacity model, the trial budgets, and the number of parallel requests (see Table~\ref{tab:zhao-parameters}). We therefore add a hybrid configuration that runs the extended model with the portfolio-overlap rule of the baseline. The hybrid and the extended configuration share the simulation logic, the calibration, and the search procedure. Both start one search per product lost on a failed edge. They differ only in the candidate pool of this search. The hybrid configuration admits firms that share a portfolio product with the failed supplier. The extended configuration admits firms that offer a product similar to the lost one. The step from the baseline to the hybrid configuration thus captures the calibration and the simulation logic. The step from the hybrid to the extended configuration captures the substitution rule alone.

Each configuration runs eight scenarios. They combine two seed types with the two spreads of the partner inertia $\kappa$ and the two removal distributions of \citet{zhao2019}. The single seed is either a hub from the top 10\% of suppliers by degree or a supplier with degree 1. Each scenario uses \num{1000} seeds. All three configurations run on the same seeds and the same imputation of the edge products (\num{24000} runs in total). The baseline keeps its original, shorter horizon (see Table~\ref{tab:zhao-parameters}).

Table~\ref{tab:exp1-arms} summarises the results. The baseline configuration never removes a firm beyond the seed. The hybrid configuration does produce cascades. With hub seeds, \qtyrange{60}{63}{\percent} of its runs remove at least one further firm. The calibration and simulation logic of the extended model make cascades possible. In the calibrated model, a candidate accepts requests only up to its free capacity, and a customer can search only as long as its inventory lasts. A lost product can therefore remain unreplaced although many firms share a product with the failed supplier. Experiment~1 does not separate which of these two mechanisms enables the cascades.

The substitution rule mainly changes how far the cascades spread. Moving from the hybrid to the extended configuration raises the share of cascading runs by only two to four percentage points. With hub seeds, however, the mean cascade grows by a factor of $4.2$ to $4.8$. The bootstrap intervals of the paired differences lie above zero in every scenario. With degree-1 seeds, the effect is small in absolute terms. With all other mechanics fixed, the substitution rule thus sets the size of the cascades. Its effect is largest where a hub fails.

\begin{table}[tbp]
  \centering
  \caption{Experiment~1: cascades of the three configurations on identical seeds ($\num{1000}$ runs per cell). Hub seeds are drawn from the 10\% of suppliers with the highest degree. Each cell gives the mean cascade size (firms removed beyond the seed) and, in parentheses, the share of runs with at least one such removal. The last column isolates the substitution rule. It gives the paired mean difference between the extended and the hybrid configuration with its \qty{95}{\percent} bootstrap interval.}
  \label{tab:exp1-arms}
  \small
  \setlength{\tabcolsep}{4pt}
  \begin{tabular}{@{}lllrrrr@{}}
    \toprule
    Seeds & $\sigma_\kappa$ & Removal & Baseline & Hybrid & Extended & Difference \\
    \midrule
    Top-10\% degree & 0.1 & normal  & 0 (0\%) & 3.37 (60\%) & 16.11 (63\%) & 12.74 [10.70, 14.74] \\
                    & 0.1 & uniform & 0 (0\%) & 3.84 (60\%) & 16.08 (64\%) & 12.24 [10.15, 14.27] \\
                    & 0.2 & normal  & 0 (0\%) & 3.78 (63\%) & 17.37 (65\%) & 13.59 [11.29, 15.86] \\
                    & 0.2 & uniform & 0 (0\%) & 3.50 (61\%) & 16.57 (64\%) & 13.06 [11.09, 15.19] \\
    \midrule
    Degree 1        & 0.1 & normal  & 0 (0\%) & 0.06 (6\%)  & 1.04 (8\%)   & 0.98 [0.36, 1.80] \\
                    & 0.1 & uniform & 0 (0\%) & 0.06 (6\%)  & 0.44 (9\%)   & 0.38 [0.11, 0.74] \\
                    & 0.2 & normal  & 0 (0\%) & 0.05 (5\%)  & 0.20 (8\%)   & 0.14 [0.04, 0.31] \\
                    & 0.2 & uniform & 0 (0\%) & 0.06 (6\%)  & 0.80 (9\%)   & 0.74 [0.26, 1.40] \\
    \bottomrule
  \end{tabular}
\end{table}

\begin{figure}[tbp]
  \centering
  \includegraphics[width=\linewidth]{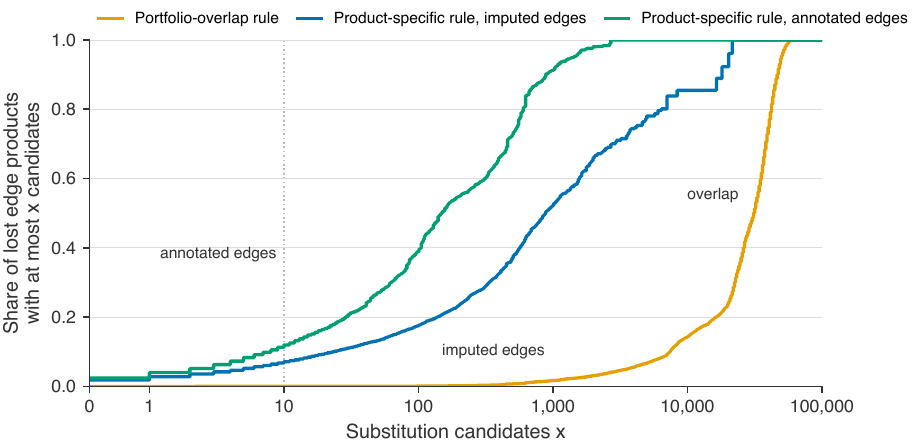}
  \caption{Substitution candidates per lost edge product under the two substitution rules. The candidates are counted on the network configuration of Experiment~1 (one Monte Carlo imputation). Each curve shows the share of lost edge products with at most $x$ candidates. The value at $x = 0$ is the share without any candidate. The portfolio-overlap rule is shown for all edges and the product-specific rule separately for imputed and annotated edges. A firm is a candidate if its portfolio or its observed deliveries contain a similar product (product-specific rule) or a product of the failed supplier (portfolio-overlap rule). Candidates exclude the failed supplier and the customer, and their free capacity is not considered. The dotted line marks ten candidates.}
  \label{fig:candidate-pools}
\end{figure}

Figure~\ref{fig:candidate-pools} shows where the two rules differ. It counts the substitution candidates of every product lost on a failed edge. The median lost product has about \num{31000} candidates under the portfolio-overlap rule and about 830 under the product-specific rule. The typical lost product has many candidates under both rules. The difference lies in the tail. Under the product-specific rule, \qty{7.2}{\percent} of the products lost on imputed edges have ten or fewer candidates and \qty{1.9}{\percent} have none. Under the portfolio-overlap rule, only \qty{0.03}{\percent} have ten or fewer. Products lost on annotated edges are scarcer still. This may partly reflect the concentration of annotated edges between Tier-1 suppliers and OEMs. It is also consistent with the hold-out result that uniform imputation understates product scarcity.

\subsubsection{Sensitivity to the imputation and the similarity threshold}\label{sec:res-sensitivity}

The data do not pin down two inputs of the extended model. The imputation decides which products sit on the unannotated edges. The similarity threshold $\tau$ decides which products count as substitutes. We vary one input at a time and repeat the hub and the degree-1 scenario on the same seeds (\num{22000} runs). For the imputation, edge products are drawn with probability proportional to $n_p^{\gamma}$. A negative $\gamma$ favours products with few substitutes. The value $\gamma = -1$ reproduces the observed edge products best in the hold-out test. We add a stronger tilt ($\gamma = -2$) and a tilt towards substitutable products ($\gamma = +1$). For the threshold, we rerun the extended configuration with values of $\tau$ between $0.65$ and $0.85$. The hybrid configuration does not use $\Sigma$ and therefore does not depend on $\tau$.

Table~\ref{tab:exp1-sensitivity} reports the factor between the mean hub cascades of the extended and the hybrid configuration. The direction of the effect holds in every setting. The extended configuration always produces larger cascades than the hybrid configuration. Only for degree-1 seeds at $\tau = 0.65$ does the bootstrap interval of the paired difference reach zero.

The size of the effect, however, depends strongly on both inputs. The factor ranges from $1.2$ to $17.7$. Imputations that place scarcer products on the edges enlarge the effect. The calibrated tilt $\gamma = -1$ raises the factor from $4.8$ to $17.6$. Only a tilt towards substitutable products ($\gamma = +1$) shrinks it to $1.2$, and the hold-out test rejects this tilt. A stricter threshold also enlarges the effect, because it leaves fewer candidates per lost product. Across the thresholds, the factor rises steadily from $1.9$ to $12.7$. The main specification lies in the lower part of the range. We regard the direction of the substitution effect as established and report its size as a range.

\begin{table}[tbp]
  \centering
  \caption{Sensitivity of Experiment~1 to the imputation tilt $\gamma$ and the similarity threshold $\tau$. Mean cascade size of the hybrid and the extended configuration with hub and degree-1 seeds ($\sigma_\kappa = 0.1$, normal removal distribution, $\num{1000}$ identical seeds per cell). The factor is the ratio of the two means for hub seeds. The hybrid configuration does not depend on $\tau$. The main specification is uniform imputation ($\gamma = 0$) with $\tau = 0.73$.}
  \label{tab:exp1-sensitivity}
  \small
  \begin{tabular}{@{}llrrrrr@{}}
    \toprule
    & & \multicolumn{3}{c}{Top-10\% degree seeds} & \multicolumn{2}{c}{Degree-1 seeds} \\
    \cmidrule(lr){3-5} \cmidrule(l){6-7}
    Input & Setting & Hybrid & Extended & Factor & Hybrid & Extended \\
    \midrule
    Imputation & $\gamma = 0$ (main)       & 3.37 & 16.11 & 4.8  & 0.06 & 1.04 \\
               & $\gamma = -2$             & 3.21 & 56.74 & 17.7 & 0.05 & 2.69 \\
               & $\gamma = -1$ (calibrated) & 2.94 & 51.79 & 17.6 & 0.08 & 3.47 \\
               & $\gamma = +1$             & 2.83 & 3.32  & 1.2  & 0.05 & 0.27 \\
    \midrule
    Threshold  & $\tau = 0.65$             & 3.37 & 6.26  & 1.9  & 0.06 & 0.41 \\
               & $\tau = 0.69$             & 3.37 & 10.98 & 3.3  & 0.06 & 0.66 \\
               & $\tau = 0.73$ (main)      & 3.37 & 16.11 & 4.8  & 0.06 & 1.04 \\
               & $\tau = 0.77$             & 3.37 & 24.84 & 7.4  & 0.06 & 1.18 \\
               & $\tau = 0.81$             & 3.37 & 34.24 & 10.2 & 0.06 & 1.44 \\
               & $\tau = 0.85$             & 3.37 & 42.78 & 12.7 & 0.06 & 1.69 \\
    \bottomrule
  \end{tabular}
\end{table}

\subsection{Experiment~2: Shock magnitude analysis and critical suppliers}\label{sec:res-exp-B}

Experiment~2 characterises how the cascade size of the extended model depends on the magnitude of the initial disruption. It also measures how fast and for how long the network loses supply capacity. The same runs show which seed firms drive the largest cascades.

We use ten seed set sizes $|S|$ between 1 and 100. For each size, we combine 500 Monte Carlo imputations of the edge products with ten random seed samples (\num{50000} runs in total). The seed firms are drawn uniformly at random from all firms with at least one customer. No targeting by degree, geography, or any other attribute is applied.

\subsubsection{Aggregate robustness curve}\label{sec:res-robust-curve}

\begin{figure}[tbp]
  \centering
  \includegraphics[width=\linewidth]{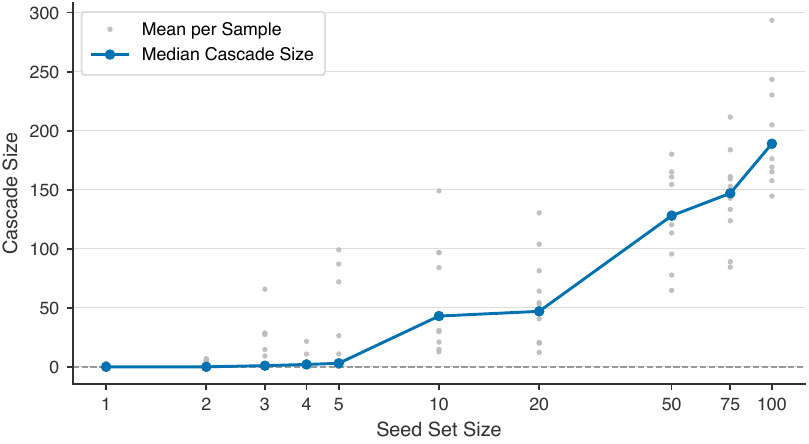}
  \caption{Robustness curve of Experiment~2. Median cascade size (firms removed beyond the seeds) as a function of the seed-set size $|S|$ (log-scaled axis). Grey dots show the ten per-seed-sample means per $|S|$. The dashed line marks the no-cascade reference. The spread of the sample means at fixed $|S|$ reflects the identity of the seeds.}
  \label{fig:robustness-e0}
\end{figure}

The robustness curve plots the median cascade size, the number of firms removed beyond the seeds, against $|S|$ (see Figure~\ref{fig:robustness-e0}). For small shocks, the identity of the seeds matters as much as their number. Up to $|S| = 5$, the median cascade stays at three removed firms or fewer. Single seed samples, however, produce mean cascades of up to 99 firms. From $|S| = 10$ onwards, the median cascade grows monotonically and reaches 189 removed firms at $|S| = 100$. At this shock size, the calibrated imputation tilt $\gamma = -1$ enlarges the mean cascade by a factor of $1.8$.

The cascades are multi-tiered. At $|S| = 100$, \qty{61}{\percent} of the cascade removals occur beyond the direct customers of the seeds (see Appendix~\ref{app:supplementary}). Random seeds are rarely hubs. Only \qty{8.8}{\percent} of them belong to the top 10\% of firms by degree. Yet \qty{94}{\percent} of the firms removed one step downstream do. Customers are reached along supply edges and thus in proportion to their connectivity. This is a supply network analogue of the friendship paradox \citep{feld1991}. The variation across seed samples is more than an order of magnitude larger than the sampling uncertainty of the imputation.

\begin{figure}[tbp]
  \centering
  \includegraphics[width=\linewidth]{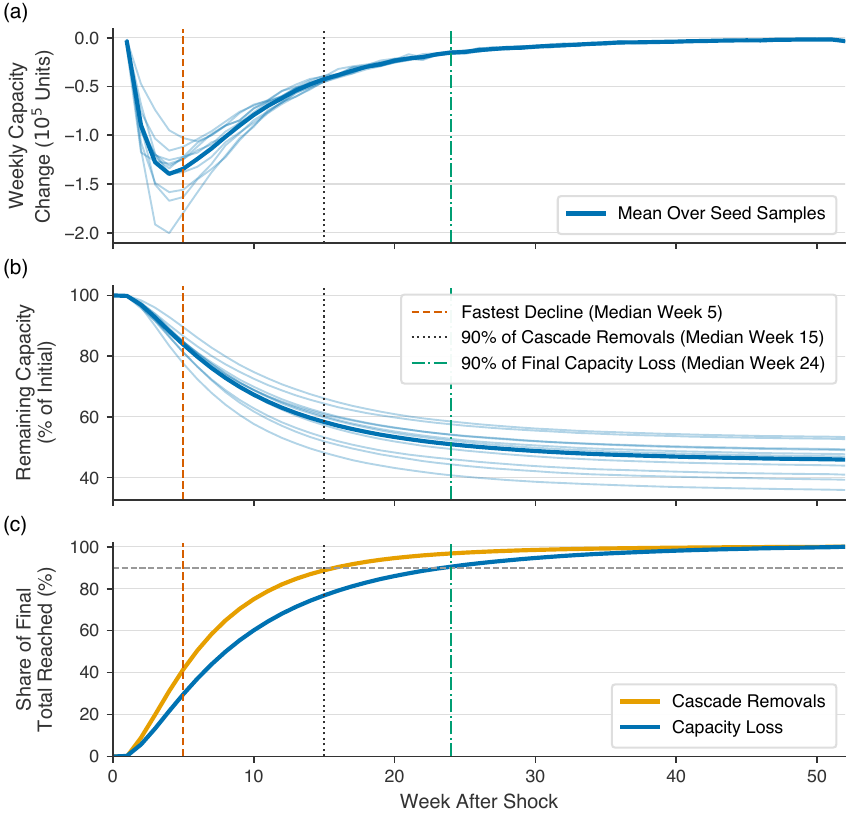}
  \caption{Supply capacity dynamics for $|S| = 100$ random seeds. \textbf{(a)}~Weekly capacity change $\Delta C(t)$. \textbf{(b)}~Remaining capacity $C(t)$ as a share of the capacity $C(0)$ before the shock. In (a) and (b), the thin lines show the mean per seed sample and the thick line the overall mean. \textbf{(c)}~Share of the final number of cascade removals and of the final capacity loss that is reached by each week, averaged over all runs. The dashed horizontal line marks \qty{90}{\percent}. In all panels, the vertical lines mark the median weeks $t_\text{fast}$, $t_{90}$, and $t_{90}^C$.}
  \label{fig:capacity-loss}
\end{figure}

Figure~\ref{fig:capacity-loss} shows how the supply capacity of the network evolves at $|S| = 100$. Three weeks characterise the dynamics. The week of fastest decline is $t_\text{fast} = \operatorname{arg\,min}_t \Delta C(t)$. The cascade duration $t_{90}$ is the first week by which \qty{90}{\percent} of the cascade removals have occurred. The week $t_{90}^C$ is the first week by which \qty{90}{\percent} of the final capacity loss has occurred. Their medians are weeks 5, 15, and 24. The loss continues long after its steepest point.

Capacity does not recover within the horizon. By week 52, the network has lost a median of \qty{54}{\percent} of its initial capacity. This follows largely from the model design. Removed firms do not return, so lost volume can only come back through substitute edges. In no run do these edges regain more than \qty{1}{\percent} of the loss. The lost capacity therefore measures the supply volume of firms that could not replace a lost input. It does not measure the time until production resumes. Capacity loss also outlasts the removals, because the late removals are fewer but larger firms. A firm removed after week 24 takes about four times as much supply volume with it as a firm removed early in the cascade. Without recovery, the network keeps losing capacity for four to six months after the shock.

\subsubsection{Critical suppliers}\label{sec:res-exp-C}

The critical-supplier analysis asks which properties make a seed firm a driver of cascades. In particular, it asks whether the largest cascades start at the hubs at the centre of the network or at less prominent firms whose loss is hard to absorb. It thereby examines the nexus supplier theory of \citet{yan2015} within the simulation on an industrial-scale network.

We reuse the $\num{50000}$ runs of Experiment~2. Five structural predictors describe each seed firm: out- and in-degree, betweenness, PageRank, and the community participation coefficient of the Louvain partition. These cover the structural indicators that \citet{yan2015} associate with the operational nexus supplier, namely degree, betweenness, and eigenvector centrality. PageRank stands in for eigenvector centrality because eigenvector centrality degenerates on a near-acyclic directed network. In a second step, we add the substitutability of the products that a firm delivers. It directly measures the non-substitutability that defines the monopolistic nexus supplier of \citet{yan2015}.

Among the classical centralities, out-degree carries the strongest signal at both the run and the firm level (see Table~\ref{tab:e0-correlations}). At the firm level, we attribute every cascade removal to the seeds that caused it by tracing the recorded disruption origins upstream. A firm's mean attributed cascade then correlates with its out-degree ($r = 0.49$). The other centralities add nothing once out-degree is controlled for.

Out-degree counts how many customers a failure exposes. It says nothing about how many of them fail. In the model, an exposed customer fails when it cannot re-source the lost products. Whether a failure propagates is thus decided edge by edge. Three properties of an edge $(u, v)$ can influence this decision. The first is the substitutability of the products on the edge. We measure it by $\ln n_p$, the logarithm of the number of firms that offer a product similar to a product $p$ on the edge. The second is the size of the customer $v$. As in the removal probability (see Equation~\ref{eq:removal}), we measure size by the degree of $v$ and use its logarithm. Larger customers are less likely to be removed. The third is the dependence of $v$ on the edge. It is the share of the supply volume that $v$ receives through this edge. For example, if 8 of the 10 product--model delivery entries that $v$ receives come from $u$, the dependence is $0.8$. Losing $u$ then raises the loss ratio $\ell(v)$ by $0.8$.

The runs show how often a failure propagates along an edge. Consider an edge $(u, v)$ that leaves a seed firm $u$. Its observed propagation rate is the share of runs seeded with $u$ in which $v$ is removed and $u$ is one of its disruption origins. Substitutability dominates this rate. On edges whose products have the fewest substitutes (bottom quartile), \qty{29.9}{\percent} of the customers fail. In the top quartile, only \qty{4.4}{\percent} do. The size of the customer and its dependence on the edge matter little.

To aggregate these rates to the firm level, we need a propagation probability for every edge downstream of a firm. The observed rate can be measured only on edges that leave seed firms. Only for these edges is the exposure clear, because a seed fails at the start of every run that includes it. Fewer than one in ten edges leave a seed firm. We therefore fit a logistic regression of the observed rate on the three edge properties. It suits the data because the rates are shares of exposed customers that fail and because its predictions remain probabilities between 0 and 1. Because substitutability dominates, the model mainly converts the number of alternative suppliers of a lost product into the probability that the customer fails. We apply it to every supply edge $(u, v) \in E_s$ to obtain its propagation probability $\phi_{uv}$. To avoid circularity, we split the seed firms randomly into five groups. The model for the firms of one group is fitted only on the edges of the other four. The observed rates on a firm's own edges thus never enter its index. On the held-out firms, the model assigns a failing customer a higher probability than a surviving one in \qty{75}{\percent} of comparisons (area under the receiver operating characteristic curve, AUC $= 0.75$). With size and dependence alone, the model does no better than chance (AUC $= 0.48$).

The substitution-aware criticality index estimates how many firms fail when a firm $u$ fails. It covers three tiers downstream: the customers of $u$, their customers, and the customers of those. A firm two tiers down fails only if the shock passes along both edges in between. Suppose that $u$ delivers a product with few substitutes to a customer $v$ and that $\phi_{uv} = 0.3$. This matches the observed rate of the bottom quartile. The failure of $u$ then removes $v$ with probability $0.3$. If $v$ fails, it removes its own customer $x$ with probability $\phi_{vx}$, say $0.2$. The failure of $u$ thus removes $x$ with probability $0.3 \times 0.2 = 0.06$. Such a probability exists for every path of up to three edges that leaves $u$. Their sum estimates the expected number of firms that the failure of $u$ removes. For example, three customers that each fail with probability $0.3$ contribute $0.9$ expected removals. The index computes this sum tier by tier. The subscript $d$ counts the tiers still to cover, so the index is $I(u) = I_3(u)$ with the recursion
\begin{equation}\label{eq:index}
  I_d(u) = \sum_{v \,:\, (u, v) \in E_s} \phi_{uv}\,\bigl(1 + I_{d-1}(v)\bigr), \qquad I_0(u) = 0.
\end{equation}
Each customer $v$ of $u$ fails with probability $\phi_{uv}$. It then counts as one removal, and $I_{d-1}(v)$ adds its onward cascade over the remaining $d - 1$ tiers.

Each further edge multiplies the contribution of a path by a probability below one. At the average propagation probability of $0.13$, a path of three edges contributes only $0.13^3 \approx 0.002$. Accordingly, extending the index to four or five edges does not improve the prediction. A firm reached along several paths counts once per path. If two customers of $u$ both supply $x$, then $x$ enters the index twice but can fail only once. The index is thus an upper bound on the expected number of removals within three tiers if every edge propagates independently.

The index differs from out-degree in two respects. It looks three tiers deep, and it gives every edge its own propagation probability. Only the second requires product data. To separate the two, we need a benchmark that looks equally deep but treats all edges alike. This benchmark is the downstream reach. Replacing every $\phi_{uv}$ with the average propagation probability $\bar{\phi}$ yields
\begin{equation}\label{eq:reach}
  R(u) = \bar{\phi}\, k^\text{out}_u + \bar{\phi}^2 w_2(u) + \bar{\phi}^3 w_3(u).
\end{equation}
Here $k^\text{out}_u$ is the out-degree of $u$, and $w_d(u)$ is the number of paths of length $d$ that leave $u$. Downstream reach depends on topology alone. Its first term is proportional to out-degree. The three measures are therefore nested. Two suppliers with the same customers and the same paths downstream illustrate what the index adds. If the first delivers products with many substitutes, its customers fail with a probability of about $0.05$. If the second delivers products with few substitutes, the probability is about $0.3$. Out-degree and downstream reach rate both suppliers equally. The index rates the second six times higher.

Table~\ref{tab:critical-index} compares the three measures as predictors of a firm's mean attributed cascade. The deeper tiers add a modest gain over out-degree. Substitutability adds a larger one. Of the 50 most damaging seed firms, the index identifies 32, downstream reach 20, and out-degree 18. The rank correlations of the three measures are almost equal, because most firms cause little damage under any of them. The gain of the index lies in finding the few firms that cause the most damage. A check on the hub seeds of the sensitivity runs confirms this ranking under every similarity threshold $\tau$ and under both imputation tilts towards scarce products (see Appendix~\ref{app:criticality-sensitivity}). Under the tilt towards substitutable products, the index performs only as well as downstream reach.

The index also holds for a shock of a different kind. We apply the propagation model of Experiment~2 unchanged to the Tohoku-inspired scenario of Experiment~3. The test covers the seed firms of this scenario that never appear as seeds in Experiment~2, so both the firms and the shock are new to the index. Of the 50 most damaging of these firms, the index identifies 30 and downstream reach 20. Downstream reach adds little over out-degree, whereas substitutability still adds a clear gain. A second test concerns the firms a regional shock reaches. Consider the firms that are never seeds but have a supplier that fails in at least one run. If every edge propagates with the same probability, the predicted removal frequency of these firms correlates with the simulated one at $r = 0.91$. With the edge-specific probabilities $\phi_{uv}$, the correlation rises to $0.97$. Network position thus largely determines where a shock lands. Substitutability adds most to predicting the damage of a failing supplier.

\begin{table}[tbp]
  \centering
  \caption{Prediction of the mean attributed cascade of a seed firm by three criticality measures. The measures are out-degree, downstream reach (see Equation~\ref{eq:reach}), and the substitution-aware index (see Equation~\ref{eq:index}). The propagation model is fitted on Experiment~2 with five-fold cross-validation over seed firms and applied unchanged to the Tohoku-inspired scenario. The evaluation covers \num{2594} seed firms of Experiment~2 and \num{4780} seed firms of the Tohoku-inspired scenario that are never seeded in Experiment~2. $r$ and $\rho$ are the Pearson and Spearman correlations with the mean attributed cascade. Top 50 counts the firms that rank among the top 50 both by the predictor and by the mean attributed cascade.}
  \label{tab:critical-index}
  \small
  \begin{tabular}{@{}lrrrrrr@{}}
    \toprule
    & \multicolumn{3}{c}{Experiment~2} & \multicolumn{3}{c}{Tohoku-inspired} \\
    \cmidrule(lr){2-4} \cmidrule(l){5-7}
    Predictor & $r$ & $\rho$ & Top 50 & $r$ & $\rho$ & Top 50 \\
    \midrule
    Out-degree                    & $0.49$ & $0.68$ & 18 & $0.55$ & $0.91$ & 17 \\
    Downstream reach & $0.64$ & $0.68$ & 20 & $0.59$ & $0.89$ & 20 \\
    Substitution-aware index & $0.83$ & $0.69$ & 32 & $0.76$ & $0.90$ & 30 \\
    \bottomrule
  \end{tabular}
\end{table}

\subsection{Experiment~3: A Tohoku-inspired regional shock}\label{sec:res-exp-D}

Experiment~3 applies a regional shock inspired by the 2011 Tohoku Earthquake to the available network. It is a scenario and does not reconstruct the 2011 event. The network represents supply relations of the early 2020s. Some of its firms, such as Hitachi Astemo and Marelli, were formed after 2011. The model also removes disrupted firms permanently, whereas the plants affected in 2011 stopped temporarily and recovered. \citet{inoue2019, carvalho2021, boehm2019, son2021} analyse the economic and network consequences of the 2011 event. Our scenario takes its spatial intensity pattern and asks how a shock of this geography and size would propagate through today's network. Almost all of the \num{9330} Japanese firms in the network are geocoded to the prefecture level. Each of them receives a failure probability based on the maximum Japan Meteorological Agency (JMA) seismic intensity observed in its prefecture during the 2011 event \citep{JMA.shindoDB}. OEMs have no location attribute in MarkLines and are therefore not part of this geocoded population. They can never be seeds, and every disruption of an OEM arises through propagation from seeded suppliers.

We test three severity scenarios with fixed failure probabilities per JMA intensity. At JMA intensities 7, 6, and 5, a firm fails with probability $0.40$, $0.20$, and $0.05$ at low severity. Medium severity raises these probabilities to $0.50$, $0.30$, and $0.10$, and high severity to $0.60$, $0.40$, and $0.15$. Each scenario runs 30 seed samples on each network configuration (\num{22500} runs in total). The outputs are the cascade size distribution per scenario and the disruption probability per firm (the share of runs in which the firm is removed). Throughout this section, a firm counts as disrupted in a run if it is removed in that run, as a seed or through the cascade. This differs from the transient Disrupted state of the model, in which a firm has failed to replace an input but is not yet removed.

\begin{figure}[tbp]
  \centering
  \includegraphics[width=0.9\linewidth]{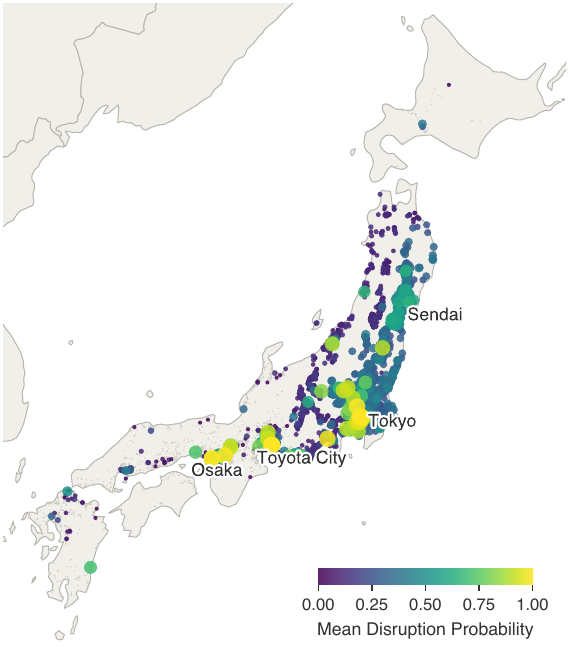}
  \caption{Geographical distribution of the disruption probability across Japan. Marker colour and area encode the firm's mean disruption probability. Vulnerable clusters appear in the Tohoku seed region around Sendai and, through propagation, in the Tokyo area and around the Tier-1 centres near Toyota City and Osaka.}
  \label{fig:geje-map}
\end{figure}

Across all runs, \num{6246} firms are disrupted at least once. Of these, \num{1253} lie outside the seed region and are reached through propagation. This union overstates the impact of a single event. A typical run disrupts a median of \num{1259} firms including the seeds, and 336 of them lie outside the seed region. The disrupted firms with a known country are located in 31 countries, mainly in Japan, the United States, China, and Germany. Figure~\ref{fig:geje-map} shows the geographical distribution of the disruption probabilities.

The disruption probability concentrates on the OEMs and the major Tier-1 suppliers (see Table~\ref{tab:geje-oem-prob}). All five primary Japanese OEMs are disrupted with a probability of at least $0.94$ at every severity. This holds even at low severity, where firms in Miyagi, the only JMA-7 prefecture, fail with a probability of only $0.40$. As the OEMs cannot be seeded, these disruptions arise entirely through propagation. The suppliers of the Japanese OEMs are concentrated in the affected prefectures. Each of the five OEMs sources from between 294 and 484 suppliers inside the seed region. In nearly every run, some critical input therefore fails. The Tier-1 suppliers are removed within the first seven weeks on average, and four of the five OEMs follow after 10 to 16 weeks. Subaru, the OEM with the highest share of suppliers inside the seed region, falls already in week 6 on average.

Among the other firms, the major Japanese Tier-1 suppliers Panasonic, Denso, Hitachi Astemo, Aisin, and Marelli have the highest disruption probabilities. Each of them is disrupted in at least \qty{99}{\percent} of the runs. They reach these probabilities mainly through propagation. Panasonic, Denso, and Aisin are never seeded, and Hitachi Astemo and Marelli only in a minority of runs. The highest-ranked non-Japanese entities are the Austrian subsidiary of Bosch, the Thai subsidiary of Yanmar, and Continental.

\begin{table}[tbp]
  \centering
  \caption{Disruption probabilities of the five primary Japanese OEMs and five major Japanese Tier-1 suppliers. The seed share is defined as the mean share of runs in which the firm is part of the seed set. The removal week is the mean week of removal over all runs in which the firm is removed.}
  \label{tab:geje-oem-prob}
  \begin{tabular}{llrrrrr}
    \toprule
    & & \multicolumn{3}{c}{Disruption probability} & & \\
    \cmidrule(lr){3-5}
    Firm & Role & Low & Medium & High & Seed share & Mean removal week \\
    \midrule
    Subaru         & OEM    & 1.000 & 1.000 & 1.000 & 0.00 & 6.4 \\
    Mazda          & OEM    & 0.993 & 0.996 & 0.995 & 0.00 & 10.6 \\
    Honda          & OEM    & 0.955 & 0.957 & 0.954 & 0.00 & 15.1 \\
    Nissan         & OEM    & 0.959 & 0.963 & 0.961 & 0.00 & 15.0 \\
    Toyota         & OEM    & 0.943 & 0.948 & 0.949 & 0.00 & 16.0 \\
    \midrule
    Panasonic      & Tier 1 & 0.997 & 1.000 & 1.000 & 0.00 & 5.7 \\
    Denso          & Tier 1 & 1.000 & 1.000 & 1.000 & 0.00 & 6.7 \\
    Hitachi Astemo & Tier 1 & 0.999 & 1.000 & 1.000 & 0.12 & 4.9 \\
    Aisin          & Tier 1 & 0.999 & 1.000 & 1.000 & 0.00 & 5.5 \\
    Marelli        & Tier 1 & 0.994 & 0.998 & 1.000 & 0.38 & 2.7 \\
    \bottomrule
  \end{tabular}
\end{table}

The temporal profile of the regional shock matches that of Experiment~2. The network loses capacity fastest in weeks 4 to 5, and \qty{90}{\percent} of the cascade removals have occurred by weeks 12 to 14. Capacity keeps falling without recovery and reaches a loss of about two thirds by week 52. Higher severity brings the fastest decline forward and makes it steeper. It also enlarges the cascades, but less than the seed set. From low to high severity, the number of seeds more than doubles, whereas the cascade beyond the seeds grows by only about \qty{40}{\percent}.

The structure of the regional propagation differs from that of the random seed cascades. As the geocoded seeds are concentrated in the Japan-led bloc, many customers lose several suppliers at once. In this scenario, \qty{34}{\percent} of the cascade removals follow the loss of two or more suppliers. In Experiment~2, the share is \qty{21}{\percent}. The cascade removals also occur earlier. The regional cascades are broad, densely cross-linked, and fast. The random seed cascades, in contrast, advance through successive failures of individual hubs in longer and deeper chains (see Figures~\ref{fig:cascade-graph-deep} and~\ref{fig:cascade-graph-tohoku}).

\section{Discussion}\label{sec:discussion}

\subsection{A regionalised but global supply network}\label{sec:disc-topology}

The network is highly modular, yet neither country nor tier alone reproduces its communities. Five large communities are each anchored in a dominant economy and extend into neighbouring supplier nations. The transatlantic bloc holds a central position among them. For the propagation experiments, the community structure has two effects. Community boundaries partly contain regional shocks. A disruption in the Japan-led bloc hits Japanese OEMs and their Thai and Chinese suppliers hard. It reaches the China-led or the transatlantic bloc only through a small set of bridge firms. At the same time, the mean shortest path length is only $3.53$, so the hubs of every community reach the other blocs within one or two further hops. A regional shock can therefore still spread globally within a few weeks. The shallow tier hierarchy, the weak clustering, and the near-zero degree assortativity describe a network in which a few hubs connect large peripheral supplier populations. In this regime, random failures are absorbed, whereas targeted hub removal collapses the giant component \citep{albert2000}. As failures propagate downstream, the critical points are the few hubs with many outgoing edges. This matches the dominance of out-degree among the centralities in Experiment~2.

\subsection{Substitutability matters for disruption propagation}\label{sec:disc-substitution}

Experiment~1 splits the gap between the baseline and the extended model into two steps. The first step, the calibrated simulation logic, makes cascades possible. Even under the portfolio-overlap rule, a hub failure removes further firms in \qtyrange{60}{63}{\percent} of the runs. Two mechanisms can explain this, the capacity limit of the candidates and the finite inventory of the customers, but Experiment~1 does not separate them. The substitution rule then determines the size of the cascades. With all other mechanics fixed, the product-specific rule makes them four to five times larger. Under imputations and thresholds that make substitutes scarcer, the factor rises to 18. Once the mechanics allow cascades, the substitution rule sets how far they spread.

Figure~\ref{fig:candidate-pools} shows the candidate pools of the two rules. For the typical lost product, both rules find many candidates. They differ for the few products with scarce substitutes. The product-specific rule leaves \qty{7.2}{\percent} of the products lost on imputed edges with ten or fewer candidates and \qty{1.9}{\percent} with none. Customers that lose such products fail most often. Under the portfolio-overlap rule, almost every firm shares some product with the failed supplier. The median lost product has about \num{31000} candidates. An overlap test that was selective among the \num{2971} firms of Mergent Horizon thus becomes permissive among the \num{76986} firms of MarkLines. The input-specificity evidence of \citet{barrot2016} and \citet{boehm2019} suggests that real substitutability is tighter still. Real substitution decisions also depend on technical specifications, qualification, tooling, logistics, and customer requirements that no network-level criterion observes. Any implementable rule, the product-specific one included, remains a proxy. The two rules of Experiment~1 mark two points on a spectrum of such proxies.

Models with a single generic good \citep{nair2011} cannot represent this constraint at all, because they do not distinguish what a supplier delivers. Substitution rules are also not scale-invariant, since candidate pools grow with network size. Small cascades under a loose rule therefore reflect the rule more than the resilience of the network. A risk assessment built on such a rule understates the damage that a product-level criterion predicts.

\subsection{Robustness curves}\label{sec:disc-robustness}

In the shock magnitude analysis of Experiment~2, small shocks are dominated by the identity of the failing firms. At $|S| \leq 5$ the median cascade stays at three removed firms or fewer, yet single seed samples produce mean cascades of up to 99 firms. For $|S| \geq 10$ the median cascade grows monotonically but less than proportionally with the shock, from 43 removed firms at $|S| = 10$ to 189 at $|S| = 100$. The cascades unfold on two timescales. The network loses capacity fastest around week~5. Only around week~15 have \qty{90}{\percent} of the removals occurred, after a slower wave has worked through the deeper tiers. Capacity keeps falling beyond that point, because the late removals are fewer but larger firms. Tracked week by week by $C(t)$ and $\Delta C(t)$, this is the ripple effect of \citet{ivanov2014} in measurable form. A risk process that reads the slowing of losses after the fastest decline as a turning point would underestimate the cumulative loss. The length of the tail is, however, an upper bound, because removed firms do not return. The timing of the acute phase follows the search and removal rules more than the inventory calibration. About nine in ten disrupted firms fail in their first search week. Fewer than one in ten edges has a budget of one week or less. Most of these firms fail because no untried candidate remains. They are removed after a median of three weeks. In the model, inventories therefore delay failures only where candidates exist.

\subsection{What makes a supplier critical}\label{sec:disc-critical}

The critical-supplier analysis decomposes the criticality of a supplier into three components. The first is exposure. A failing supplier hits only its own customers, and every cascade starts with them. Their number, the out-degree, therefore predicts the damage of a supplier better than any other classical centrality. For the same reason, a run causes more damage the more customers its seeds have in total. The second component, downstream reach, adds the customers of the customers, so a supplier whose customers are hubs themselves causes larger cascades than its out-degree suggests. Systemic-risk indices such as that of \citet{diem2022} also trace how a failure propagates downstream. Like exposure, downstream reach follows from topology alone. The third component, substitutability, decides whether an exposed customer fails. Topology cannot reveal it, yet it dominates propagation along individual edges. It also gives the largest gain in identifying the most damaging suppliers. Of the top 50, the index identifies 32 and downstream reach 20. Under the regional shock, the gain over out-degree comes mostly from substitutability. The firms this shock reaches are mostly predictable from network position. For an exposed firm, the removal frequency predicted from topology alone correlates with the simulated one at $r = 0.91$. With substitutability, the correlation rises to $0.97$.

For the nexus supplier theory of \citet{yan2015}, which locates criticality in network position and not in internal capabilities, the result is mixed. Position enters through out-degree and downstream reach. Betweenness adds nothing beyond out-degree, because a cascade spreads from a failed supplier to all of its customers at once and not along brokered short paths. The most damaging suppliers stand out through the substitutability of what they deliver. This supports the monopolistic nexus supplier of \citet{yan2015}, whose criticality lies in not being substitutable. With product data, this property can be measured directly instead of being proxied through betweenness in the surrounding industry network. The concentration of cascade risk in a few such firms resembles the finding of \citet{diem2022} for the Hungarian VAT network.

The index approximates the simulation but does not replace it. It misses 18 of the 50 most damaging suppliers, and extending it beyond three tiers does not close this gap. The index treats every edge in isolation and assumes independent propagation. The simulation additionally resolves three interactions. Searching customers compete for the same free capacity. Inventories limit how long a customer can search. In a regional shock, the losses from many failing suppliers add up at their customers. Only the simulation yields the timing of a cascade and the capacity it destroys. The practical use of the index therefore lies in screening. A supplier-risk process with product data on its supply relationships can rank suppliers by how hard their products are to replace and how far a failure would reach downstream. The simulation then shows how a cascade from the flagged suppliers would unfold.

The attack tolerance picture of \citet{albert2000} also changes in a product-annotated supply network. The damage a hub causes depends on its degree and on how hard its products are to re-source. Randomness, moreover, protects less than in the classical picture. Only \qty{8.8}{\percent} of random seeds are top-10\%-degree firms, but \qty{94}{\percent} of the firms they remove one step downstream are. A random local failure thus acts as a hub removal one tier below the initial disruption.

\subsection{Regional vulnerabilities: a Tohoku-inspired scenario}\label{sec:disc-geje}

The community structure provides a natural frame for the Tohoku-inspired scenario. Most firms in the prefectures with the highest JMA intensities belong to the Japan-led bloc. This bloc also contains the affected OEMs and their supplier base in Thailand and China. The strongest external connection of this bloc is the transatlantic bloc. Of the \num{1253} disrupted firms outside the seed region, \qty{42}{\percent} belong to the Japan-led bloc itself, \qty{25}{\percent} to the transatlantic bloc, and \qty{21}{\percent} to the China-led bloc.

The disruption probabilities of the five primary Japanese OEMs lie between $0.94$ and $1.00$ and vary by at most $0.006$ across the scenarios (see Table~\ref{tab:geje-oem-prob}). This reflects their exposure to 294 to 484 suppliers each in the seed region. A firm counts as disrupted when it is removed after failing to replace a lost input. The probability therefore measures the risk of an unresolved input loss. It does not measure the risk of a permanent closure. The temporal structure matches that of Experiment~2. Capacity falls fastest four to five weeks after the shock, and \qty{90}{\percent} of the removals have occurred by weeks 12 to 14. Capacity keeps falling long after that, and \qty{90}{\percent} of the final loss is reached only around weeks 20 to 22. Higher severity brings the fastest decline forward and makes it steeper. Yet although the seed mass more than doubles, the week of fastest decline moves by only one week. The search and removal rules thus set the timescale more than the seed mass. The concentration of the seeds in one community, however, changes the structure of the propagation. Joint supplier losses are more frequent than under random seeding. They precede \qty{34}{\percent} of the cascade removals. Under random seeding, the share is \qty{21}{\percent}. They raise the loss ratio $\ell(f)$ of the affected customers and produce broad, fast, and shallow cascades instead of long chains of hub failures.

Historical observations provide context for the scenario, but they are not a comparable outcome measure. In 2011, the earthquake halted vehicle production at all major Japanese OEMs. Japanese vehicle output in April 2011 was an estimated \qty{80}{\percent} below the pre-quake rate \citep{canis2011}, and the OEMs resumed production from mid-April at reduced volumes \citep{Honda.2011, Toyota.2011, Nissan.2012}. The model, by contrast, removes disrupted firms permanently, has no recovery dynamics, and runs on the network of the early 2020s. We therefore do not compare the simulated timing or the affected firms with the historical record. Such a comparison would require firm-level records of which suppliers halted production and for how long.

\subsection{Strengths and limitations}\label{sec:disc-strengths}

The study has three strengths. First, the model combines candidate selection by the specific lost product with supplier rewiring. Other models resolve substitution by sector \citep{inoue2019} or by product \citep{inoue2025, amico2024} but keep the set of suppliers fixed. The model of \citet{zhao2019} lets customers rewire but selects candidates by firm portfolio. Experiment~1 holds all other mechanics fixed. The product-level rule then makes the cascades four to five times larger. Second, agent-based models with supplier rewiring have so far run on networks such as the \num{2971}-firm Mergent Horizon network of \citet{zhao2019}. The MarkLines network is more than an order of magnitude larger. It also carries product annotations on its edges. Third, the Tohoku-inspired scenario applies a spatially correlated shock derived from observed seismic intensities to a model with supplier rewiring. Prior agent-based models with supplier rewiring, such as that of \citet{zhao2019}, use stylised random or top-degree disruption scenarios. \citet{inoue2019} apply the observed damage of the 2011 earthquake to nearly one million firms but keep supply relationships fixed.

These strengths are subject to four limitations. First, only $\qty{6.3}{\percent}$ of MarkLines edges carry an explicit product annotation. The remaining $\qty{93.7}{\percent}$ are imputed via Monte Carlo simulation. This imputation is the dominant non-stochastic source of model uncertainty. A hold-out test on the annotated edges shows that uniform imputation understates product scarcity. The sensitivity runs show that the direction of the substitution effect does not depend on the imputation, but its size does. The factor between the hybrid and the extended configuration ranges from $1.2$ to $17.7$. Absolute cascade sizes are best read as one point within a range. Second, the product-specific rule rests on the semantic similarity of product names as a proxy for technical substitutability. The product taxonomy and observed dual sourcing bound its errors but cannot eliminate them. The size of the substitution effect also grows with the threshold $\tau$. The same proxy enters the criticality index. The index and the simulation both count substitutes with it, so the index fits the simulated cascades partly by construction. It shows how much of the simulated criticality substitutability explains. Whether real criticality follows the same pattern can only be tested against observed disruptions. Third, the network snapshot is static. The model has no firm entry or exit, no recovery dynamics, and no mitigation channels such as expedited logistics. This limits analyses with a time horizon beyond one quarter. In particular, the capacity decline beyond the acute phase holds only in the absence of recovery. Depending on the tiers of the edge, the inventory buffers last three to ten weeks on average. If failed firms resumed production within this time, part of the late removals would not occur and the tail would be shorter. In addition, a customer without untried candidates becomes Disrupted at once, even if its inventory would last longer. The model brings these failures forward by up to the budget of the edge. With supplier recovery, some of them would not occur. Fourth, attributes are systematically missing in the MarkLines dataset. OEM entities carry no country or location attribute, so they can never be part of a geocoded seed set. Furthermore, $\qty{59}{\percent}$ of the disrupted firms outside the seed region of the Tohoku-inspired scenario lack a populated country attribute. Such gaps add to the missing firms and links that bias network statistics in commercial datasets \citep{bacilieri2026}. They propagate into any analysis that conditions on the affected attributes.

\section{Conclusions}\label{sec:conclusions}

We asked how portfolio-overlap and lost-product-specific candidate-selection rules change cascades under shared supplier-search and rewiring conditions. On the product-annotated MarkLines network of the global automotive industry, the choice of rule changes the size of the cascades. With all other mechanics fixed, the product-specific substitution rule makes the cascades after a hub failure four to five times larger than the portfolio-overlap rule. The direction of this effect holds in every sensitivity setting. Its size ranges from 1.2 to 18 and depends on how scarce substitutes are. Because only \qty{6}{\percent} of the edges carry observed products, the data constrain this scarcity only partly.

The effect arises from the few lost products with scarce substitutes. On a network of this size, almost every firm shares some product with a failed supplier. A portfolio-overlap rule therefore finds substitutes even for products that a product-level criterion cannot replace. A rule that is selective on a small network becomes permissive on an industry-scale network. Under such a rule, a large network appears more robust than a product-level criterion suggests.

Network topology largely determines which firms a shock reaches. Only product data show whether these firms can re-source the lost products. Within the simulations, substitutability also decides which suppliers are critical. Of the 50 most damaging suppliers under random disruptions, downstream reach identifies 20 and an index that adds substitutability identifies 32. The index carries over without refitting to the regional scenario modelled on the 2011 Tohoku Earthquake. This supports the view of \citet{yan2015} that a monopolistic nexus supplier is critical because it cannot be replaced. Once calibrated by simulation, the index can screen suppliers in supplier-risk processes. The course of a cascade and the capacity it destroys still require the simulation. Risk assessments of supply networks should therefore account for product-level substitutability. This requires product data on supply relationships.

Four directions for further research follow. First, substitutability criteria should be validated against observed supplier switches, qualification records, or procurement data. This would give propagation models a firmer empirical basis. Second, a tier-dependent reactivation process would add recovery dynamics and allow post-shock policies to be compared. Third, scenarios modelled on qualitatively different shocks, such as the 2021 Renesas semiconductor fire or the COVID-19 lockdowns, would test how far the findings generalise. Fourth, the time-resolved disruption probabilities of the Tohoku-inspired scenario quantify correlated upstream risk for a focal firm. They could enter risk-aware production scheduling or procurement models.

\backmatter

\bmhead{List of abbreviations}
ARIO~--~Adaptive Regional Input--Output (model),
AUC~--~Area Under the Receiver Operating Characteristic Curve,
B2B~--~Business-to-Business,
BFS~--~Breadth-First Search,
CCDF~--~Complementary Cumulative Distribution Function,
DAG~--~Directed Acyclic Graph,
GDP~--~Gross Domestic Product,
HPC~--~High-Performance Computing,
JMA~--~Japan Meteorological Agency,
KDE~--~Kernel Density Estimate,
KS~--~Kolmogorov--Smirnov,
LCC~--~Largest (Weakly) Connected Component,
LLM~--~Large Language Model,
OEM~--~Original Equipment Manufacturer,
RFQ~--~Request for Quote,
VAT~--~Value-Added Tax.

\section*{Declarations}

\subsection*{Availability of data and materials}
The MarkLines data underlying this study are proprietary and cannot be published. They are available from MarkLines Co.,~Ltd.\ under licence \citep{marklinescoltd2026}, and restrictions apply to their redistribution. The simulation and analysis code, including the configurations and random seeds of all experiments, is available as follows.
\begin{itemize}
  \item Project name: edge-product-disruption-sim
  \item Project home page: \url{https://github.com/benjaminrolf/edge-product-disruption-sim}
  \item Archived version: v1.1.0 at Zenodo, \url{https://doi.org/10.5281/zenodo.23701335} \citep{Rolf2026.code}
  \item Operating system: platform independent
  \item Programming language: Python 3.13
  \item Other requirements: the Python packages pinned in the repository, Graphviz for the layout of Figure~\ref{fig:network-overview}
  \item Licence: MIT
  \item Restrictions to use by non-academics: none for the code. The MarkLines data require a licence.
\end{itemize}
The simulation outputs supporting the conclusions of this article are available from the corresponding author on request in anonymised form with firm identifiers removed.

\subsection*{Competing interests}
The authors declare that they have no competing interests.

\subsection*{Funding}
This work was supported by the German Federal Ministry of Research, Technology and Space (BMFTR, formerly Federal Ministry of Education and Research, BMBF) under grant no. 16DKWN092A. The funder had no role in the design of the study, the analysis and interpretation of the data, or the writing of the manuscript.

\subsection*{Authors' contributions}
BR conceptualised the study, developed the methodology and the software, performed the formal analysis and the visualisation, wrote the original draft, and acquired the funding. HI contributed to the conceptualisation, supervised the work, curated the data, and provided resources. SL contributed to the conceptualisation and supervised the work.

\subsection*{Acknowledgements}
The authors used a large language model (Claude, Anthropic) to assist with language editing, restructuring of the manuscript text, and the development of analysis scripts. All content was reviewed and verified by the authors, who take full responsibility for the work.

\begin{appendices}

\section{Dataset construction and coverage}\label{app:data}

\begin{figure}[tbp]
  \centering
  \includegraphics[width=\linewidth]{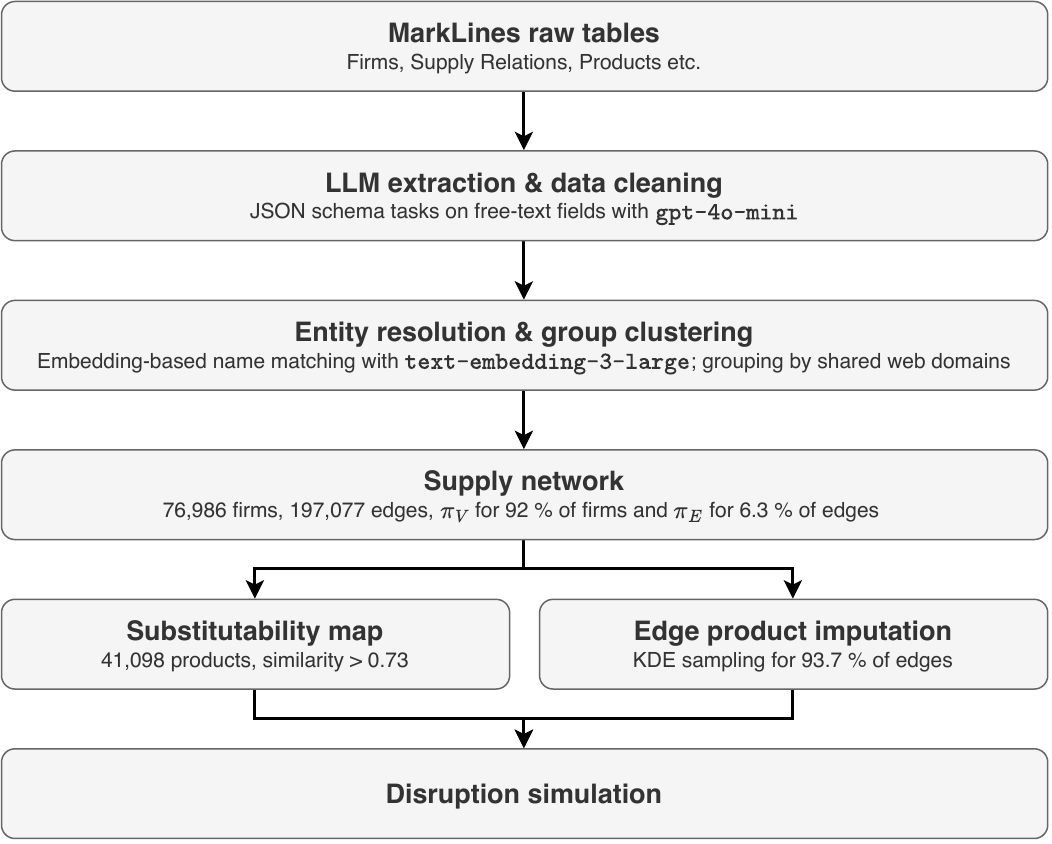}
  \caption{Data and simulation pipeline. Raw MarkLines tables are cleaned by an LLM, reconciled via embeddings, clustered into firm groups, and assembled into the supply network. The product substitutability map $\Sigma$ and the Monte Carlo edge-product imputation feed the disruption simulation.}
  \label{fig:pipeline}
\end{figure}

MarkLines does not provide a polished supply network dataset. Its records mix plant-level entities (e.g.\ Toyota Motor Kyushu, Miyata Plant) with their corporate parents (e.g.\ Toyota Motor Corporation) within the same table. The relevant information is spread across separate tables for firms, supply relations, products, shareholders, and branches, and many attributes are stored as free text. The reconstruction therefore proceeds in three stages. Figure~\ref{fig:pipeline} summarises them together with the two steps that prepare the network for the simulation.

The first stage turns free-text fields into structured attributes. It uses the \texttt{gpt-4o-mini} model of OpenAI through the Batch API with structured outputs. All outputs feed deterministic downstream processing.

The second stage resolves entities and clusters firm groups. Records belonging to the same firm group are linked in a similarity graph. Its edges represent shared URL domains and name strings recovered from the firm record and from the parsed shareholder relations. The connected components of this graph then define the group clusters. Firm names that appear across the supply, partner, and shareholder tables are then reconciled with the firm detail records. The reconciliation uses a cosine-similarity threshold of $0.73$ on embeddings from the \texttt{text-embedding-3-large} model of OpenAI. The threshold conservatively merges legal form variations while keeping distinct subsidiaries of the same parent as separate entities. Entity resolution itself relies on embedding similarity rather than generative output.

The third stage assembles the consolidated records into the supply network with its product maps $\pi_V$ and $\pi_E$. The substitutability map $\Sigma$ and the Monte Carlo edge-product imputation then prepare the network for the simulation.

The network that results from these stages differs from the data on which the baseline model was developed. The relevant reference point is the Mergent Horizon dataset. \citet{zhao2019} built their agent-based model on it. Its properties frame the baseline comparison of Experiment~1. Table~\ref{tab:marklines-vs-mergent} summarises the size and topology statistics of both datasets. MarkLines is more than an order of magnitude larger than Mergent Horizon along every dimension relevant to propagation experiments. It has roughly 26 times more nodes, 21 times more edges, and a denser high-degree tail (top-10\% degree range $17$--$\num{3804}$ vs.\ $15$--$312$ in Mergent). Despite the size difference, the undirected diameter is comparable ($10$ vs.\ $13$). In a small-world network, the diameter grows only logarithmically with $n$.

\begin{table}[tbp]
  \centering
  \caption{Comparison of MarkLines and Mergent Horizon. n/a: not available.}
  \label{tab:marklines-vs-mergent}
  \begin{tabular}{lll}
    \toprule
    Metric & MarkLines & Mergent Horizon \\
    \midrule
    Nodes (total) & \num{76986} & \num{2971} \\
    Nodes (LCC) & \num{32668} & n/a \\
    Edges (total) & \num{197077} & \num{9535} \\
    Edges (LCC, undirected) & \num{197021} & n/a \\
    Diameter (undirected) & 10 (LCC) & 13 \\
    Top-10\% degree range & 17--\num{3804} & 15--312 \\
    Edge-level product information & yes (\qty{6.3}{\percent} of edges) & no \\
    \bottomrule
  \end{tabular}
\end{table}

This comparison motivates the experimental design. In a network with 26 times as many firms and a heavier high-degree tail, the typical disrupted supplier has far more alternative suppliers. We therefore expect the portfolio-overlap rule of \citet{zhao2019} to permit substitution too readily and to underpredict propagation. The product-specific rule requires edge-level product attributes, and only MarkLines carries them.

\section{Extended topology analysis}\label{app:topology}
\setcounter{figure}{0}\setcounter{table}{0}

The topology is characterised in two steps. First, the degree distributions are fitted against heavy-tailed reference distributions. Second, the standard structural measures are reported in three groups, namely small-world structure and tier hierarchy, community structure, and assortativity and connectivity.

\subsection{Degree distribution}

\begin{figure}[tbp]
  \centering
  \includegraphics[width=\linewidth]{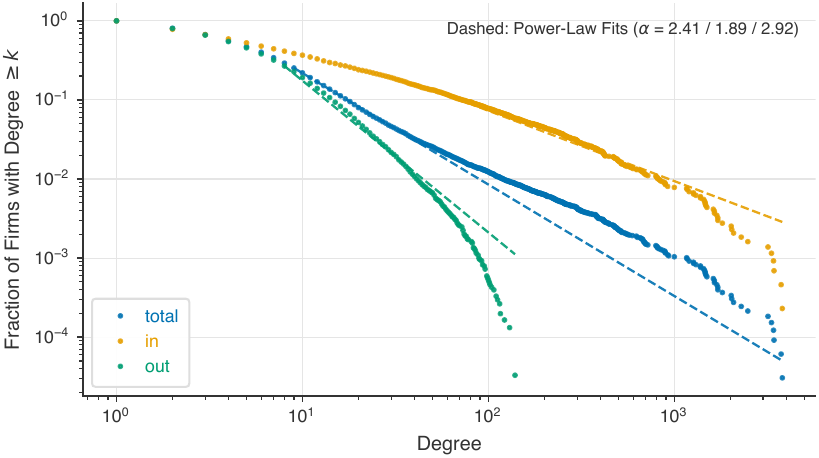}
  \caption{Degree CCDFs with maximum-likelihood power-law fits. Total, in-, and out-degree distributions of the MarkLines supply network on log--log axes. Dashed lines show the discrete maximum-likelihood power-law fits above their respective $x_\text{min}$. The power law is preferred only for the total degree. Log-normal fits are preferred for the in- and out-degree distributions.}
  \label{fig:degree-ccdf}
\end{figure}

The total, in-, and out-degree distributions are heavy-tailed (CCDFs in Figure~\ref{fig:degree-ccdf}). We fit a discrete power law $P(k) \propto k^{-\alpha}$ via maximum likelihood and compare it against exponential and log-normal alternatives using the normalised log-likelihood ratio test as applied by \citet{broido2019}. The estimated tail exponents are $\alpha_\text{total} = 2.41$, $\alpha_\text{in} = 1.89$, and $\alpha_\text{out} = 2.92$. All three distributions reject the exponential alternative ($p < 10^{-3}$), so the tails are unambiguously heavy. For the FactSet supply chain dataset, \citet{bacilieri2026} estimate exponents of $1.72$ for the in-degree and $2.36$ for the out-degree. MarkLines shows the same ordering. The heavier in-degree tail reflects a small number of large OEMs that absorb extreme numbers of incoming edges. No supplier dominates the customer side of the network to the same extent, hence the steeper out-degree tail ($\alpha_\text{out} = 2.92$). The tests prefer the power law only for the total degree. For the in- and out-degree distributions separately, a log-normal fit is statistically preferred ($p \leq 0.007$). \citet{broido2019} likewise find that log-normal distributions fit most real networks as well as or better than power laws. We therefore use ``heavy-tailed'' throughout the paper and treat the power-law exponents as descriptive summaries of the tails rather than as evidence of strict scale-free dynamics.

\subsection{Small-world structure and tier hierarchy}

Clustering is measured on the directed LCC with the average clustering coefficient for directed networks of \citet{fagiolo2007}. This coefficient counts the directed triangles around each firm. The observed value $C_\text{cl} = 0.028$ is small in absolute terms but is $1.44$ times the mean of a directed degree-sequence-preserving random baseline (configuration model, $50$ draws) and significant ($z = 27.6$). Triangles form more often than chance predicts, for example when a supplier serves both a Tier-1 firm and the OEM that this Tier-1 firm supplies. This excess depends on the null model. Under an undirected configuration model on the same LCC, the comparison inverts. The observed $C_\text{cl} = 0.056$ then lies below the random expectation of $0.103$. The network is nearly acyclic (see the bow-tie decomposition below) and therefore admits far fewer triangles than an undirected random graph with the same degree sequence. The directed null model preserves the supplier and customer roles of each firm and is therefore the appropriate reference for a supply network. The mean shortest path length on the undirected LCC is $L = 3.53$, well below the LCC diameter of $10$ and below the random-graph expectation $L_\text{random} = \ln N / \ln \langle k \rangle = 4.17$. The resulting small-world index $\sigma = (C_\text{cl} / C_\text{cl,random}) / (L / L_\text{random}) = 1.70$ exceeds one, but it reflects short paths combined with a weak excess of directed triangles rather than the strongly clustered regime of \citet{watts1998}.

\subsection{Community structure}

The Louvain partition is comparable in strength to the highly modular Airbus supply network of \citet{brintrup2017}. Two structural baselines show that no obvious firm attribute fully explains the partition. MarkLines records no country for \num{1852} firms of the LCC. They comprise its \num{1183} OEMs and 669 other firms without customers. On the remaining \num{30816} firms, a pure country partition reaches $Q_\text{country} = 0.40$, whereas the Louvain partition reaches $Q = 0.51$. A pure tier partition is anti-modular ($Q_\text{tier} = -0.259$). This follows by construction, because supply edges run between tiers rather than within them. The five blocs are unevenly interconnected. The transatlantic bloc has the largest number of inter-bloc edges (\num{28291}, or \qty{38}{\percent} of the edges incident to it). It is also the largest external partner of each of the other four blocs. The strongest single inter-bloc connection links the Japan-led and the transatlantic bloc with \num{11146} edges. The Japan-led and the China-led bloc share \num{4353} edges. The degree of regional closure differs between blocs. The share of external edges is \qty{22}{\percent} for the China-led bloc, the most self-contained, \qty{35}{\percent} for the Japan-led and \qty{40}{\percent} for the India-led bloc. The small Korea-led bloc has \qty{63}{\percent} external edges, mainly to the transatlantic and Japan-led blocs.

\subsection{Assortativity and connectivity}

All four directional variants of the degree assortativity of \citet{newman2002} are close to zero (between $-0.053$ and $+0.002$). High-degree OEMs source from large peripheral supplier populations rather than from one another. A bow-tie decomposition \citep{broder2000} confirms a near-DAG signature. Only $68$ firms ($\qty{0.09}{\percent}$ of nodes) form the largest strongly connected component. In the canonical bow-tie of the web graph, this component holds around \qty{28}{\percent} of the nodes. The periphery is dominated by $\num{44244}$ isolated firms with no recorded supply ties. As reciprocal supply loops are rare, a disrupted customer can seldom re-source through cyclic supply relations. Substitutes must come laterally, from competing suppliers that offer similar products.

\section{Validation of the imputation and the substitution rule}\label{app:validation}
\setcounter{figure}{0}\setcounter{table}{0}

\subsection{Hold-out test of the imputation}

For each of the \num{12411} annotated edges, we hide the observed products and keep their number $k$. We then draw $k$ products from the supplier's portfolio $\pi_V(u)$, 20 times per edge, either uniformly or with probability proportional to $n_p^{\gamma}$. For every observed and every imputed product, we count the substitution candidates as the simulation does. These are the firms other than the supplier and the customer of the edge whose portfolio or observed deliveries contain a similar product. Table~\ref{tab:holdout} compares the two candidate distributions. The tilt $\gamma$ is chosen to minimise the Kolmogorov--Smirnov distance between the log candidate counts of imputed and observed products. In a five-fold cross-validation over edges, every fold selects $\gamma = -1$. Its test distances of $0.08$ to $0.10$ are far below those of the uniform draw ($0.33$ to $0.35$). Two properties of the annotated edges limit how far this result transfers to the other edges. First, the annotated edges come almost entirely from the last stage of the supply chain. Of the annotated edges, \qty{99.3}{\percent} end at a firm without customers of its own. Of the unannotated edges, \qty{77.9}{\percent} do. Second, only \qty{27}{\percent} of the observed product names also occur in the portfolio vocabulary. These products have a median of 528 candidates, and only \qty{0.7}{\percent} of them have ten or fewer. The remaining products have a median of 100 candidates, and \qty{16.1}{\percent} have ten or fewer. The scarcity of the observed products lies mainly in names that no portfolio contains. The imputation cannot draw such names. Part of this scarcity may therefore reflect differences in naming rather than a lack of substitutes. For both reasons, we keep the uniform draw as the main specification and treat the tilted draws as sensitivity cases.

\begin{table}[htbp]
  \centering
  \caption{Hold-out test of the edge-product imputation on the \num{12411} annotated edges. Candidates are the firms offering a product similar to the given one. The Kolmogorov--Smirnov (KS) distance compares the log candidate counts of imputed and observed products. Recall is the share of observed products with a similar product among those imputed on the same edge.}
  \label{tab:holdout}
  \small
  \begin{tabular}{@{}lrrrrr@{}}
    \toprule
    & \multicolumn{3}{c}{Candidates} & & \\
    \cmidrule(lr){2-4}
    Products & Median & $\leq 10$ & None & KS & Recall \\
    \midrule
    Observed                        & 147  & \qty{12.0}{\percent} & \qty{2.5}{\percent} & --     & -- \\
    Imputed, $\gamma = +1$          & 5999 & \qty{0.5}{\percent}  & \qty{0.1}{\percent} & $0.65$ & \qty{12.9}{\percent} \\
    Imputed, $\gamma = 0$ (uniform) & 623  & \qty{5.4}{\percent}  & \qty{1.2}{\percent} & $0.34$ & \qty{18.2}{\percent} \\
    Imputed, $\gamma = -0.5$        & 323  & \qty{14.9}{\percent} & \qty{4.3}{\percent} & $0.14$ & \qty{19.8}{\percent} \\
    Imputed, $\gamma = -1$          & 196  & \qty{19.2}{\percent} & \qty{5.7}{\percent} & $0.08$ & \qty{20.3}{\percent} \\
    Imputed, $\gamma = -2$          & 146  & \qty{21.0}{\percent} & \qty{6.5}{\percent} & $0.11$ & \qty{20.7}{\percent} \\
    \bottomrule
  \end{tabular}
\end{table}

\subsection{Plausibility checks of the substitution map}

The two checks cannot validate technical substitutability. They do, however, bound the errors of the similarity proxy in both directions. The taxonomy check estimates how often the proxy matches products that cannot replace each other. The dual-sourcing check estimates how often it misses real substitutes. The taxonomy check uses the 13 product families and 117 product groups into which MarkLines classifies its \num{963} standard products. Among the $\Sigma$ pairs of two different standard products, the share that falls into different families is a lower bound for implausible matches. An example is a seat part matched with an engine part. The dual-sourcing check uses the products that an OEM sources from two suppliers for the same vehicle model and model year (\num{3282} directed supplier pairs). Each supplier of such a pair is a revealed substitute for the other. The recall is the share of pairs in which the rule lists the second supplier as a candidate when the first fails. The recall is reported once with the portfolios alone and once with the observed deliveries $\pi_D$. The second variant counts only deliveries to customers other than the one of the pair. Table~\ref{tab:sigma-checks} reports both checks across the thresholds of the sensitivity runs. A stricter threshold reduces the cross-family matches and the candidate pools, but it also misses more revealed substitutes. Under the portfolio-overlap rule, the dual-sourcing recall is \qty{91.8}{\percent}.

\begin{table}[htbp]
  \centering
  \caption{Plausibility checks of the substitution map $\Sigma$ across the similarity threshold $\tau$. Pairs are $\Sigma$ pairs of two different standard products. Recall is the share of revealed dual-sourcing substitutes that the product-specific rule lists as candidates. It is given for the portfolios alone and with the observed deliveries. The last column gives the median number of candidates of the dual-sourced products.}
  \label{tab:sigma-checks}
  \small
  \begin{tabular}{@{}lrrrrrr@{}}
    \toprule
    & & \multicolumn{2}{c}{Taxonomy} & \multicolumn{2}{c}{Dual-sourcing recall} & \\
    \cmidrule(lr){3-4} \cmidrule(lr){5-6}
    $\tau$ & Pairs & Same group & Other family & Portfolio & + deliveries & Candidates \\
    \midrule
    0.65 & 1590 & \qty{65.3}{\percent} & \qty{21.5}{\percent} & \qty{79.9}{\percent} & \qty{90.6}{\percent} & 473 \\
    0.69 & 910  & \qty{70.3}{\percent} & \qty{17.7}{\percent} & \qty{70.0}{\percent} & \qty{87.4}{\percent} & 336 \\
    0.73 & 510  & \qty{72.8}{\percent} & \qty{16.5}{\percent} & \qty{64.4}{\percent} & \qty{86.1}{\percent} & 146 \\
    0.77 & 297  & \qty{74.8}{\percent} & \qty{16.5}{\percent} & \qty{56.6}{\percent} & \qty{83.8}{\percent} & 44 \\
    0.81 & 172  & \qty{80.8}{\percent} & \qty{13.4}{\percent} & \qty{50.0}{\percent} & \qty{81.5}{\percent} & 13 \\
    0.85 & 74   & \qty{82.4}{\percent} & \qty{13.5}{\percent} & \qty{47.3}{\percent} & \qty{80.3}{\percent} & 5 \\
    \bottomrule
  \end{tabular}
\end{table}

\section{Supplementary results}\label{app:supplementary}
\setcounter{figure}{0}\setcounter{table}{0}

\subsection{Propagation depth and disruption graphs}\label{app:depth}

\begin{figure}[htbp]
  \centering
  \includegraphics[width=\linewidth]{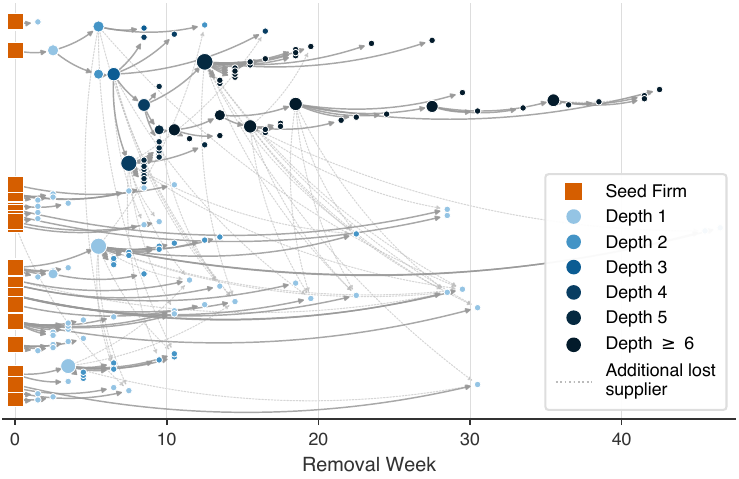}
  \caption{Time-resolved disruption graph of the deepest random-seed cascade of Experiment~2. The run has $|S| = 75$ seeds, $141$ cascade removals, and a propagation depth of $11$. Squares mark the $23$ seeds that triggered at least one removal. Circles mark cascade removals, coloured by propagation depth and sized by the number of removals they trigger directly. Solid arrows run from the primary failed supplier to the customer it triggered. Dotted arrows mark further lost suppliers.}
  \label{fig:cascade-graph-deep}
\end{figure}

The cascades are multi-tiered. The propagation depth is the longest directed chain of removed firms downstream of any seed. All but one of the $\num{5000}$ runs with $|S| = 100$ reach a depth of at least two. The median longest chain spans four edges, and $\qty{61}{\percent}$ of all cascade removals occur at depth ${\geq}2$, i.e. beyond the direct customers of the seed set. Figure~\ref{fig:cascade-graph-deep} shows the run with the deepest removal chain of the campaign ($|S| = 75$, propagation depth $11$) as a disruption graph. It shows the characteristic geometry of these random seed cascades. As the degree distribution is heavy-tailed, a uniformly drawn seed is rarely a hub. Across the campaign, the seeds have a median degree of $4$, and only \qty{8.8}{\percent} of them fall into the top-10\% degree class (degree 17 or more). The firms removed one step downstream are a different population. They are reached along supply edges, and thus in proportion to their connectivity. This is a supply network analogue of the friendship paradox \citep{feld1991}: the removals at depth 1 have a median degree of $384$, and \qty{94}{\percent} of them are top-10\%-degree firms. Once such a hub fails, the disruption accelerates because its many customers are exposed at once, and the resulting chains stretch the cascade over more than \qty{40}{\week}.

\begin{figure}[htbp]
  \centering
  \includegraphics[width=\linewidth]{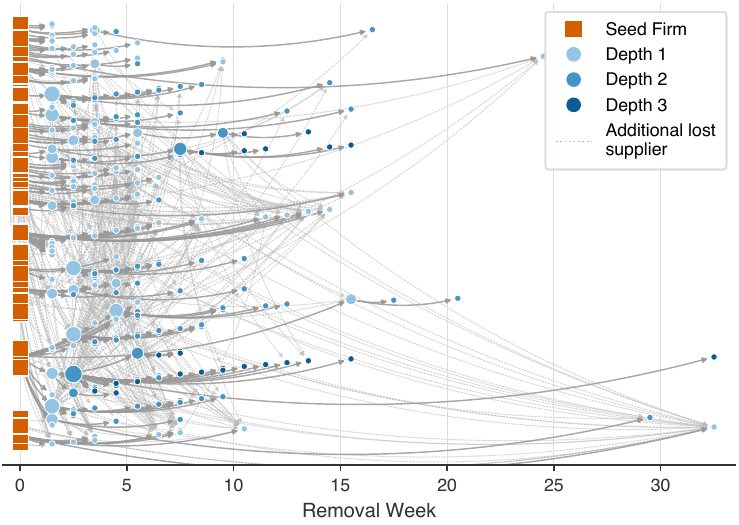}
  \caption{Time-resolved disruption graph of a low-severity run of the Tohoku-inspired scenario. The run has $506$ seeds and $249$ cascade removals. Squares mark the $70$ seeds that triggered at least one removal. Markers and arrows follow Figure~\ref{fig:cascade-graph-deep}. Of the removed customers, $105$ (\qty{42}{\percent}) lost two or more suppliers.}
  \label{fig:cascade-graph-tohoku}
\end{figure}

The structure of the regional propagation differs from the random seed cascades of Experiment~2, as the disruption graphs in Figures~\ref{fig:cascade-graph-deep} and~\ref{fig:cascade-graph-tohoku} illustrate. As the geocoded seeds are concentrated in the Japan-led bloc, many customers lose several suppliers at once. Across the \num{22500} Tohoku runs, \qty{34}{\percent} of the \num{9345392} cascade removals follow the loss of two or more distinct suppliers. In Experiment~2, the share is \qty{21}{\percent}. In the run shown in Figure~\ref{fig:cascade-graph-tohoku} the share even reaches \qty{42}{\percent}. In the removal mechanism of the model, the weekly removal probability increases with the lost share of inbound supply $\ell(f)$, so joint supplier losses lead to faster removals. The median cascade removal occurs in week $4.5$ in the Tohoku-inspired scenario versus week $6.5$ in Experiment~2 (90th percentile: week $11.5$ vs.\ $17.5$). The regional disruption graph is consequently broad and densely cross-linked but shallow. In the depicted run, its propagation depth is at most 3. The random seed cascades of Experiment~2, in contrast, advance through successive failures of individual hub suppliers and unfold as longer and deeper chains.

\subsection{Centralities and edge properties of the seed firms}\label{app:centralities}

At the run level, the sums of out-degree, PageRank, and the participation coefficient over the seed set all correlate with the total removal count at about $r = 0.9$ (see Table~\ref{tab:e0-correlations}). This mainly reflects the seed mass. Runs with more seeds remove more firms, and the count includes the seeds themselves. Once the seed count is held fixed, only the mean out-degree of the seeds carries a clear signal (partial $r = 0.20$).

At the firm level, we attribute each cascade removal to the seeds that caused it by tracing the recorded disruption origins upstream. Where a removal has several origins, we split it equally among them. For each of the $\num{2594}$ unique seed firms, we then average the attributed removals over the runs in which the firm is a seed. All \num{3194056} cascade removals across the \num{50000} runs are traceable to at least one seed. Under this attribution, out-degree is the dominant structural predictor of a firm's mean attributed cascade (Pearson $r = 0.49$). No other centrality adds signal once out-degree is controlled for (partial $|r| \leq 0.07$).

The propagation model of the critical-supplier analysis measures the substitutability of an edge by $\ln n_p$. As in the simulation, $n_p$ counts the firms whose portfolio or observed deliveries contain a product similar to $p$. On annotated edges we use the scarcest observed product. On imputed edges the imputation draws from the supplier's portfolio, so we average $\ln n_p$ over this portfolio. For the dependence, imputed edges carry the mean volume of the annotated edges.

\begin{table}[htbp]
  \centering
  \caption{Pearson correlations of seed-firm centralities with the outcomes of Experiment~2. At the run level, the sum over the seed set is correlated with the total removal count. The partial correlation relates the seed-set mean to the cascade size after adjusting both for $|S|$. At the firm level, a seed firm's value is correlated with its mean attributed cascade.}
  \label{tab:e0-correlations}
  \begin{tabular}{lrrr}
    \toprule
    & \multicolumn{2}{c}{Run level} & Firm level \\
    \cmidrule(lr){2-3} \cmidrule(lr){4-4}
    Predictor & $r$ (sum) & Partial $r$ (mean) & $r$ \\
    \midrule
    Out-degree                      & $0.91$ & $0.20$  & $0.49$ \\
    In-degree                       & $0.54$ & $0.06$  & $0.13$ \\
    Betweenness centrality          & $0.40$ & $0.03$  & $0.11$ \\
    PageRank                        & $0.90$ & $0.10$  & $0.12$ \\
    Community participation coefficient (Louvain)       & $0.90$ & $0.04$  & $0.13$ \\
    \bottomrule
  \end{tabular}
\end{table}

\subsection{Sensitivity of the criticality comparison}\label{app:criticality-sensitivity}

The imputation tilt $\gamma$ and the similarity threshold $\tau$ change both the simulated cascades and the substitutability $\ln n_p$ of the edges. We therefore repeat the comparison of Table~\ref{tab:critical-index} for every setting of Table~\ref{tab:exp1-sensitivity}. Experiment~2 has no runs under these settings. We use the hub seeds of the Experiment~1 sensitivity runs instead. Each setting simulates the same \num{1000} runs. Every run has a single seed from the top 10\% of suppliers by degree. These runs cover 649 distinct seed firms. In each setting, $n_p$ is counted with the similarity map of its threshold. On imputed edges, the portfolio mean of $\ln n_p$ is weighted with $n_p^\gamma$, as in the imputation. The propagation model is refitted for each setting with the same five-fold cross-validation over seed firms. No sensitivity runs exist for the Tohoku-inspired scenario.

Table~\ref{tab:critical-sensitivity} shows the results. Under the main calibration, the hub seeds reproduce the ranking of Table~\ref{tab:critical-index}. Out-degree performs worse than in Experiment~2, because all seeds are hubs. The index ranks first under every threshold and under both tilts towards scarce products. Under the tilt towards substitutable products ($\gamma = +1$), it performs only as well as downstream reach. This tilt places mostly substitutable products on the imputed edges. Few customers then fail for lack of substitutes, and the cascades are small (see Table~\ref{tab:exp1-sensitivity}). Substitutability no longer predicts propagation, and the cross-validated AUC of the propagation model drops to $0.49$. The index thus adds information where substitutes are scarce enough to stop cascades.

\begin{table}[htbp]
  \centering
  \caption{Sensitivity of the criticality comparison of Table~\ref{tab:critical-index} to the imputation and the similarity threshold. The settings vary the imputation tilt $\gamma$ and the similarity threshold $\tau$. Hub seeds of the Experiment~1 sensitivity runs ($\num{1000}$ identical runs per setting, 649 distinct seed firms). $r$ and $\rho$ are the Pearson and Spearman correlations with the mean attributed cascade. Top 50 counts the firms that rank among the top 50 both by the predictor and by the mean attributed cascade.}
  \label{tab:critical-sensitivity}
  \small
  \setlength{\tabcolsep}{4pt}
  \begin{tabular}{@{}lrrrrrrrrr@{}}
    \toprule
    & \multicolumn{3}{c}{Out-degree} & \multicolumn{3}{c}{Downstream reach} & \multicolumn{3}{c}{Index} \\
    \cmidrule(lr){2-4} \cmidrule(lr){5-7} \cmidrule(l){8-10}
    Setting & $r$ & $\rho$ & Top 50 & $r$ & $\rho$ & Top 50 & $r$ & $\rho$ & Top 50 \\
    \midrule
    $\gamma = 0$ (main)        & $0.27$ & $0.36$ & 10 & $0.51$ & $0.37$ & 20 & $0.68$ & $0.48$ & 31 \\
    $\gamma = -2$              & $0.21$ & $0.32$ & 6  & $0.54$ & $0.32$ & 23 & $0.77$ & $0.59$ & 36 \\
    $\gamma = -1$ (calibrated) & $0.24$ & $0.33$ & 6  & $0.57$ & $0.36$ & 23 & $0.75$ & $0.60$ & 33 \\
    $\gamma = +1$              & $0.28$ & $0.37$ & 14 & $0.44$ & $0.34$ & 22 & $0.44$ & $0.34$ & 21 \\
    \midrule
    $\tau = 0.65$              & $0.32$ & $0.32$ & 15 & $0.46$ & $0.33$ & 19 & $0.60$ & $0.36$ & 22 \\
    $\tau = 0.69$              & $0.26$ & $0.32$ & 9  & $0.45$ & $0.36$ & 19 & $0.57$ & $0.42$ & 26 \\
    $\tau = 0.73$ (main)       & $0.27$ & $0.36$ & 10 & $0.51$ & $0.37$ & 20 & $0.68$ & $0.48$ & 31 \\
    $\tau = 0.77$              & $0.26$ & $0.33$ & 10 & $0.59$ & $0.37$ & 22 & $0.70$ & $0.51$ & 31 \\
    $\tau = 0.81$              & $0.22$ & $0.28$ & 6  & $0.59$ & $0.40$ & 24 & $0.70$ & $0.55$ & 31 \\
    $\tau = 0.85$              & $0.19$ & $0.29$ & 6  & $0.61$ & $0.46$ & 22 & $0.73$ & $0.60$ & 35 \\
    \bottomrule
  \end{tabular}
\end{table}

\end{appendices}

\bibliography{references}

\end{document}